\pdfoutput=1
\documentclass[a4paper,fleqn,usenatbib]{mn2e}

\usepackage{times}

\usepackage[T1]{fontenc}
\usepackage{ae,aecompl}

\usepackage{graphicx}	
\usepackage{amsmath}	
\usepackage{amssymb}	
\usepackage{todonotes}
\usepackage{bm}
\usepackage{subfig}
\usepackage{courier}
\usepackage[T1]{fontenc}
\usepackage{aecompl}

\long\def\symbolfootnote[#1]#2{\begingroup%
\def\thefootnote{\fnsymbol{footnote}}\footnote[#1]{#2}\endgroup} 

\title[Dynamical Constraints on Outer Planets in Super-Earth Systems]{Dynamical Constraints on Outer Planets in Super-Earth Systems}

\author[M. J. Read et al.]{
Matthew J. Read$^{1}$\thanks{E-mail: mjr201@ast.cam.ac.uk}
and Mark C. Wyatt$^{1}$
\\
$^{1}$Institute of Astronomy, University of Cambridge, Madingley Road, Cambridge CB3 0HA \\
}

\date{Accepted XXX. Received YYY; in original form ZZZ}

\pubyear{2015}

\begin{document}
\label{firstpage}
\pagerange{\pageref{firstpage}--\pageref{lastpage}}
\maketitle

\begin{abstract}
This paper considers secular interactions within multi-planet systems. In particular we consider dynamical evolution of known planetary systems resulting from an additional hypothetical planet on an eccentric orbit. We start with an analytical study of a general two-planet system, showing that a planet on an elliptical orbit transfers all of its eccentricity to an initially circular planet if the two planets have comparable orbital angular momenta. Application to the single Super-Earth system HD38858 shows that an additional hypothetical planet below current radial velocity (RV) constraints with {\textit{Msini}}=3-10M$_\oplus$, semi-major axis 1-10au and eccentricity 0.2-0.8 is unlikely to be present from the eccentricity that would be excited in the known planet (albeit cyclically). However, additional planets in proximity to the known planet could stabilise the system against secular perturbations from outer planets. Moreover these additional planets can have an {\textit{Msini}} below RV sensitivity and still affect their neighbours. For example, application to the two Super-Earth system 61Vir shows that an additional hypothetical planet cannot excite high eccentricities in the known planets, unless its mass and orbit lie in a restricted area of parameter space. Inner planets in HD38858 below RV sensitivity would also modify conclusions above about excluded parameter space. This suggests that it may be possible to infer the presence of additional stabilising planets in systems with an eccentric outer planet and an inner planet on an otherwise suspiciously circular orbit. This reinforces the point that the full complement of planets in a system is needed to assess its dynamical state.

\end{abstract}




\section{Introduction}
\begin{table*}
	\begin{center}
		\hspace*{-0.5cm}
		\begin{tabular}{ | c | c | c | c | c |c| } 
			\hline
			Planet & a (au) & {\textit{Msini}} (M$_\oplus$) & e & Period (days) &  $\omega$ (degrees) \\ 
			\hline
			HD38858b & 0.642 $\pm$ 0.002 & 12 $\pm$ 2 & - & 198 $\pm$ 1 & - \\ 
			61Vir-b & 0.050201 $\pm$ 0.000005 & 5.1 $\pm$ 0.5 & 0.12 $\pm$ 0.11 & 4.22 & 105 $\pm$ 54\\ 
			61Vir-c & 0.2175 $\pm$ 0.0001 & 18.2 $\pm$ 1.1 & 0.14 $\pm$ 0.06 & 38.02 & 341 $\pm$ 38\\ 
			\hline
		\end{tabular}
		\caption{Orbital parameters for the planets of HD38858 and 61Vir given in Kennedy et al. 2015 from Marmier et al. in prep and Vogt et al. 2010.}
		\label{tab:61Vir}
	\end{center}
\end{table*}
Since the discovery of the first stellar multi-planetary system $\nu$ Andromedae (\cite{1999ApJ...526..916B}), the number of current detections stands at 487\footnote{\label{note1}exoplanet database, exoplanet.eu (\cite{2011A&A...532A..79S})}. Compared with single systems, multiplicity introduces dynamical interactions between planets, leading to an evolution of orbital elements. Whether dynamics can be used to probe the formation mechanisms of exoplanets is an on-going topic of research, however it is possible that evidence of early interactions may be imprinted on the orbital elements of planets in current observations. An often quoted example is the relatively large eccentricities seen across the exoplanet population (e.g. \cite{2006ApJ...646..505B}; \cite{2011arXiv1109.2497M}; \cite{2012MNRAS.425..757K}) which cannot be explained by migratory formation models alone, as coupling between the planet and protoplanetary disk promotes the circularisation of orbits (\cite{1993ARA&A..31..129L}). One dynamical process believed to account for this is post-formation gravitational planet-planet scattering brought on by a dynamical instability after the protoplanetary disk has dispersed (\cite{1996Sci...274..954R}; \cite{2008ApJ...686..603J}; \cite{2009ApJ...699L..88R}), with it being likely that secular interactions further evolve eccentricities on long timescales (\cite{1996Sci...274..954R}; \cite{1997ApJ...477..781L}; \cite{2001AJ....121.1736Y}; \cite{2004AJ....128..869Z}). 

Our understanding of different scenarios of dynamical interactions in multi-planetary systems during formation and beyond can also have implications for additional planets. For example, a dearth of low mass, close-in orbit planets is expected in areas of Hot-Jupiter migration (\cite{2015ApJ...808...14M}). The spacing of planets has also been attributed to transfer of angular momentum between inner low mass and outer high mass gas giant planets (\cite{1997A&A...317L..75L}). The number of Jupiter sized objects scattered out to wide orbits detectable via direct imaging has also been predicted to be related to the number and radial distribution of close in planets (\cite{2010EAS....42..419V}). Thus the presence or absence of additional planets within a given system may give some insight into constraints on evolutionary history. 

\begin{figure}
	\includegraphics[width = \linewidth]{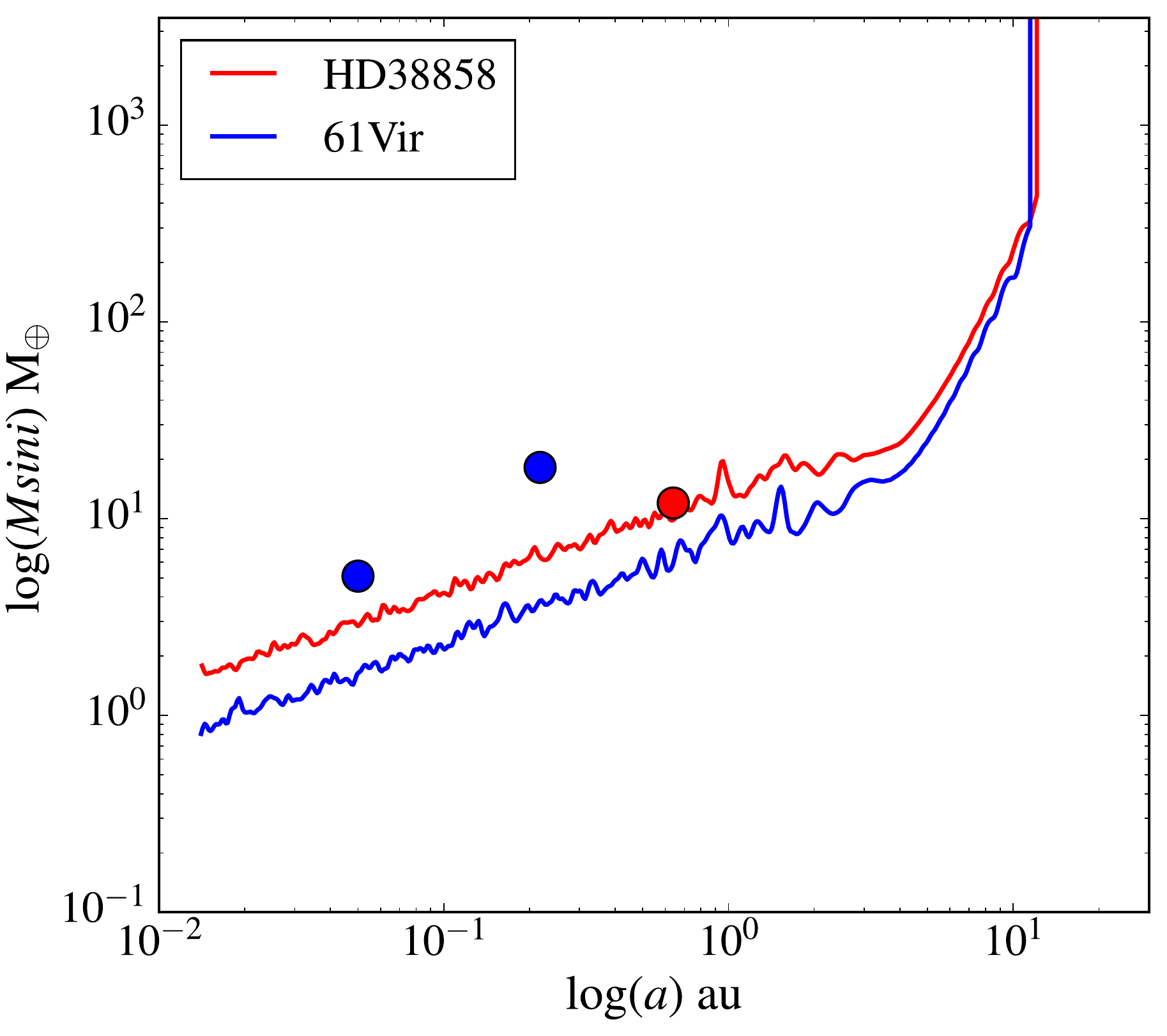}
	\caption{Super -Earth systems HD38858 (red) and 61Vir (blue). The respective planets of each system are represented by circles, alongside the current HARPS RV sensitivity. Sensitivity is quantified through identifying whether the signal generated by simulated planets is above the observational threshold, which is approximately that expected for 1m/s which has a semi-major axis dependence of $\sim a^{1/2}$ up to $\sim$8au due to the survey being active for $\sim$10yrs, not long enough to detect full periods from planets on wider orbits.}
	\label{fig:HD3861Vir}
\end{figure}

Our full interpretation of the dynamical state of known exoplanet systems is impeded by the fact that we have only partial knowledge of the planets that are present. Currently the HARPS radial velocity (RV) survey (\cite{2003Msngr.114...20M}) is one of the most sensitive to planet detection, measuring Doppler shifts down to as little as $\sim$1m/s, with detections resulting in an estimate that $\gtrsim$50\% of solar-type stars harbour at least one planet (\cite{2011arXiv1109.2497M}). The sensitivity of HARPS is shown in Figure \ref{fig:HD3861Vir}, when applied to the Super-Earth systems HD38858 and 61Vir (blue and red lines respectively, \cite{2003Msngr.114...20M}; \cite{2011arXiv1109.2497M}; \cite{2015MNRAS.449.3121K}). Since these systems are discussed in detail later in this paper we summarize the orbital parameters of the contained planets in Table \ref{tab:61Vir} (\cite{2010ApJ...708.1366V}; \cite{2015MNRAS.449.3121K}; Marmier et al. in prep). HARPS can only detect objects above the RV sensitivity, and provides no information on the existence or otherwise of objects below, as associated Doppler shifting would be too small or the orbital period too long. That is, Earth type planets on close in, $\sim$0.1au, orbits or any sized object outside of $\sim$10au could be present in these systems without contradicting current observations. 

\indent We aim to understand how the dynamics of exoplanet systems are affected by the presence of extra hypothetical planets. This work focuses on systems currently known to have 1 or 2 planets, and in particular considers the long term secular interactions with an additional planet on an eccentric orbit. We subsequently explore the possibility of placing constraints (in addition to those placed by HARPS) on such hypothetical planets, by identifying where these objects would be expected to induce eccentricities in known planets that are significantly larger than observed values. We also consider whether these eccentricities induced in confirmed planets would be large enough to cause a potential scattering/collisional event between neighbouring planets. We focus application on a general 2 planet system and then to the specific planetary systems HD38858 and 61Vir. The close proximity of HD38858 and 61Vir and associated HARPS RV measurements provides the best chance for future follow-up studies to detect additional planets, for which this work can be used to guide. These systems also each have the benefit of an imaged debris disk at $\sim$30au (\cite{2012MNRAS.424.1206W}; \cite{2015MNRAS.449.3121K}), which is assumed to provide an outer constraint on the orbits of hypothetical planets and an estimate for orbital inclination of the planetary system. Understanding the dynamics of these systems may also go some way to explaining the abundance of Super - Earths in exoplanetary systems, as it is estimated that as many as 30-50\% of G-type stars host such objects (\cite{2013ApJS..204...24B}). 

In this paper we first describe dynamical interactions that exist in planetary systems in \S\ref{sec:Dist}, focusing on how secular interactions can be used to describe the evolution of eccentricity and how this can be transferred between planets. We apply this to a general 2-body system in \S\ref{sec:2bod} and to HD38858b interacting with an additional object in \S\ref{sec:HD38858} to investigate the dynamical evolution of a 2 (planet)-body system subject to secular perturbations and discuss any constraints that can be placed on hypothetical planets. Extension of this analysis to a 3-body system is discussed through application to 61Vir secularly interacting with an additional planet in \S\ref{sec:61Vir}. We discuss a method of using secular theory to infer the presence of planets in \S\ref{sec:stable}, before comparing dynamical evolution made using secular theory with N-body simulations in \S\ref{sec:simulation}. A final conclusion of this work is then given in \S\ref{sec:conc}.

\section{The Disturbing Function}
\label{sec:Dist}
The dynamical evolution of planetary systems is driven by interweaving planet-planet interactions in addition to the more obvious interaction with the host star. Perturbations in semi-major axis ({\textit{a}}), eccentricity ({\textit{e}}), longitude of pericentre ($\varpi$), inclination ({\textit{I}}), longitude of ascending node ({\textit{$\Omega$}}) and mean longitude ($\lambda$) caused by one planet's gravitational potential on another can be described through use of the disturbing function. By expanding the disturbing function as an infinite sum of these orbital elements, the associated evolution of each respective element can be determined. The expansion contains terms that can be categorised into three types of interactions: secular, resonant and fast, which we discuss below.

\subsection{Secular Interactions}
\label{subsec:secular}
Secular components of the disturbing function represent long term interactions, with typical timescales being at least the orbital period over the planet to star mass ratio. The associated terms in the disturbing function are those independent of mean longitude, which leads to semi-major axis and therefore energy being constant. That is, secular interactions result in changes to ($e,\varpi,I,\Omega$), and angular momentum can be exchanged but not energy. If eccentricities and inclinations remain low then the corresponding evolution of each respective element can be calculated by taking the disturbing function expansion to second order in eccentricity and inclination which is known as Laplace - Lagrange theory (e.g. \cite{1999ssd..book.....M}). In this situation one finds the evolution of $e$ and $\varpi$ to be completely decoupled from the evolution of $I$ and $\Omega$. Here we discuss the derivation for the evolution of eccentricity and longitude of pericentre.

 Consider a system of $N$ planets in which planet $j$ has an eccentricity and longitude of pericentre $e_j$ and $\varpi_j$ respectively which can be combined to give a complex eccentricity $z_j = e_je^{i\varpi_j}$ so that $e_j(t)$ = |$z_j(t)$|. If $\bm{z}$ = [$z_1,z_2,..z_N$] then Laplace - Lagrange theory gives the time evolution of eccentricities of planets around a star of mass $M_c$ in the form
\begin{equation}
\bm{\dot{z}} = iA\bm{z},
\label{eq:zdot}
\end{equation}
where $i$ refers to the imaginary unit here only and $A$ is a matrix with elements given by 
\begin{equation}
A_{ji} = -\frac{1}{4}n_{j}\frac{m_i}{M_c + m_j}\alpha_{ji}\tilde{\alpha}_{ji}b^{(2)}_{3/2}(\alpha_{ji}) \hspace{5mm}(j \neq i),
\end{equation}
\begin{equation}
A_{jj} = \frac{1}{4}n_{j}\sum^{N}_{i=1, i\neq j}\frac{m_i}{M_c + m_j}\alpha_{ji}\tilde{\alpha}_{ji}b^{(1)}_{3/2}(\alpha_{ji}),
\label{eq:A}
\end{equation}
$j$ and $i$ are integers which refer to corresponding planets, $n_j$ the mean motion where $n_j^2a_j^3 \approx GM_{c}$, $\alpha_{ji}$ = $\tilde{\alpha}_{ji}$ = $a_j/a_i$ for $a_j$ < $a_i$ and $\alpha_{ji}$ = $a_i/a_j$ and $\tilde{\alpha}_{ji}$ = 1 for $a_j$ > $a_i$.
Laplace coefficients are given by
\begin{equation}
b^{(\nu)}_{s}(\alpha) = \frac{1}{\pi} \int^{2\pi}_{0}\frac{\cos(\nu x)dx}{(1 - 2\alpha\cos(x) + \alpha^2)^s}, \hspace{5mm} \alpha < 1.
\label{eq:laplace}
\end{equation} 

As eq (\ref{eq:zdot}) can be represented by a set of linear differential equations with constant coefficients, it can be solved as an $N$x$N$ eigenfrequency problem. The solution represents the evolution of $\bm{z}$ through a superposition of sinusoids associated with each eigenfrequency $g_i$ of the matrix $A$
\begin{equation}
z_j(t) = \sum^N_{i=1} \bm{e}_{ji} e^{i(g_it +\beta_i)},
\label{eq:zevo}
\end{equation}
where $\bm{e}_{ji}$ includes the eigenvectors of $A$ and initial conditions of the system and $\beta_i$ is an initial phase term. Each planet therefore imposes an eccentricity variation in the other planets on a timescale of $\sim$1/$g_i$. 
 
Description of the secular evolution by Laplace - Lagrange theory begins to break down as eccentricities are increased, specifically above $e$$\sim$0.2, where it is necessary to include higher order terms in the disturbing function. Beyond $e$$\sim$0.66 the expansion of the disturbing function no longer converges (known as the Sundman criterion), so it is not clear to what extent the second order eccentricity terms will capture any aspect of the secular interaction, although it may give an approximation. High inclinations also add further complications to the secular solution as the evolution of inclinations starts to be coupled with the evolution of eccentricity, resulting in qualitatively different behaviour. 

The evolution of a massless test particle can also be fully described within Laplace - Lagrange theory, with the full mathematical derivation given in \cite{1999ssd..book.....M}. This finds that the test particle eccentricity precesses in a circle around forced elements that are imposed upon it by the secular solution of the planets. Locations where the particle's rate of precession $A$ is equal to one of the eigenfrequencies introduces a singularity into the secular solution, causing forced elements to become infinite. Seeing as $A$ is dependant on semi-major axis, this suggests test particles will have infinite eccentricity at discrete positions, known as secular resonances. These resonances are implicit to Laplace-Lagrange theory however, since including higher order terms causes $g_i$ to vary in time, allowing for resonances to move, even to the point where they overlap. Ensuing chaotic motions can subsequently take place on timescales as long as $\sim$Gyrs (\cite{2011ApJ...739...31L}). Planets that are very low in mass compared with other planets may be expected to behave in a similar way to a test particle.

\subsection{Resonant effects}
\label{subsec:res}
Resonant terms in the disturbing function become important when the orbital periods of two bodies ($1$ and $2$) are spaced by an integer ratio. These mean motion resonances (MMRs) occur when the ratio of their semi-major axes is given by
\begin{equation}
\frac{a_2}{a_1} = \left(\frac{p}{q}\right)^{2/3},
\label{eq:aMMR}
\end{equation}
where $p$ and $q$ are integers. MMRs have finite widths and the region close to a planet is densely packed with overlapping first order resonances inducing chaotic motion in test particles on timescales of $\sim$kyr -- Myrs (\cite{2012MNRAS.419.3074M}; \cite{2015ApJ...799...41M}). The semi-major axis from the planet, inside which instability is predicted ($\Delta a$) is given by
\begin{equation}
\Delta a = 1.5a\mu^{2/7},
\label{eq:chaoswid}
\end{equation} 
where $a$ is the semi-major axis of the planet and $\mu$ is equal to the planet to star mass ratio, $m$/$M_{c}$ (\cite{1980AJ.....85.1122W}; \cite{1989Icar...82..402D}; \cite{2014A&A...563A..72F}).

\subsection{Scattering Interactions}
\label{subsec:scat}
Fast interactions in the disturbing function are often assumed to cause rapid changes in orbital elements that average to zero over single periods and therefore associated terms in the disturbing function can be neglected. In the event of scattering or collisional interactions between two bodies this is not the case, which can occur when their orbits are separated by a small enough distance. A rough estimate for when such close encounters can occur can be made by considering whether the bodies approach each other within some number of mutual Hill radii. For an object of mass $m_1$ and semi-major axis $a_1$ and a neighbouring object of mass $m_2$ and semi-major axis $a_2$ the mutual Hill radii $R_H$, using the assumption of low eccentricity, is given by 
\begin{equation}
R_H = \frac{1}{2}\left(\frac{m_{1} + m_{2}}{3M_{c}}\right)^{1/3}(a_1 + a_2),
\label{eq:mutualhill}
\end{equation}
where $M_c$ is the mass of the central star. Defining mutual Hill radii in this way does not properly account for interactions between objects on highly eccentric orbits, for which a modified form of eq \eqref{eq:mutualhill} outlined in \cite{2014MNRAS.443.2541P} can be used to express $R_H$ in terms of apocentre/pericentre distances.

\begin{figure*}
	\centering
	\includegraphics[width=0.42\linewidth]{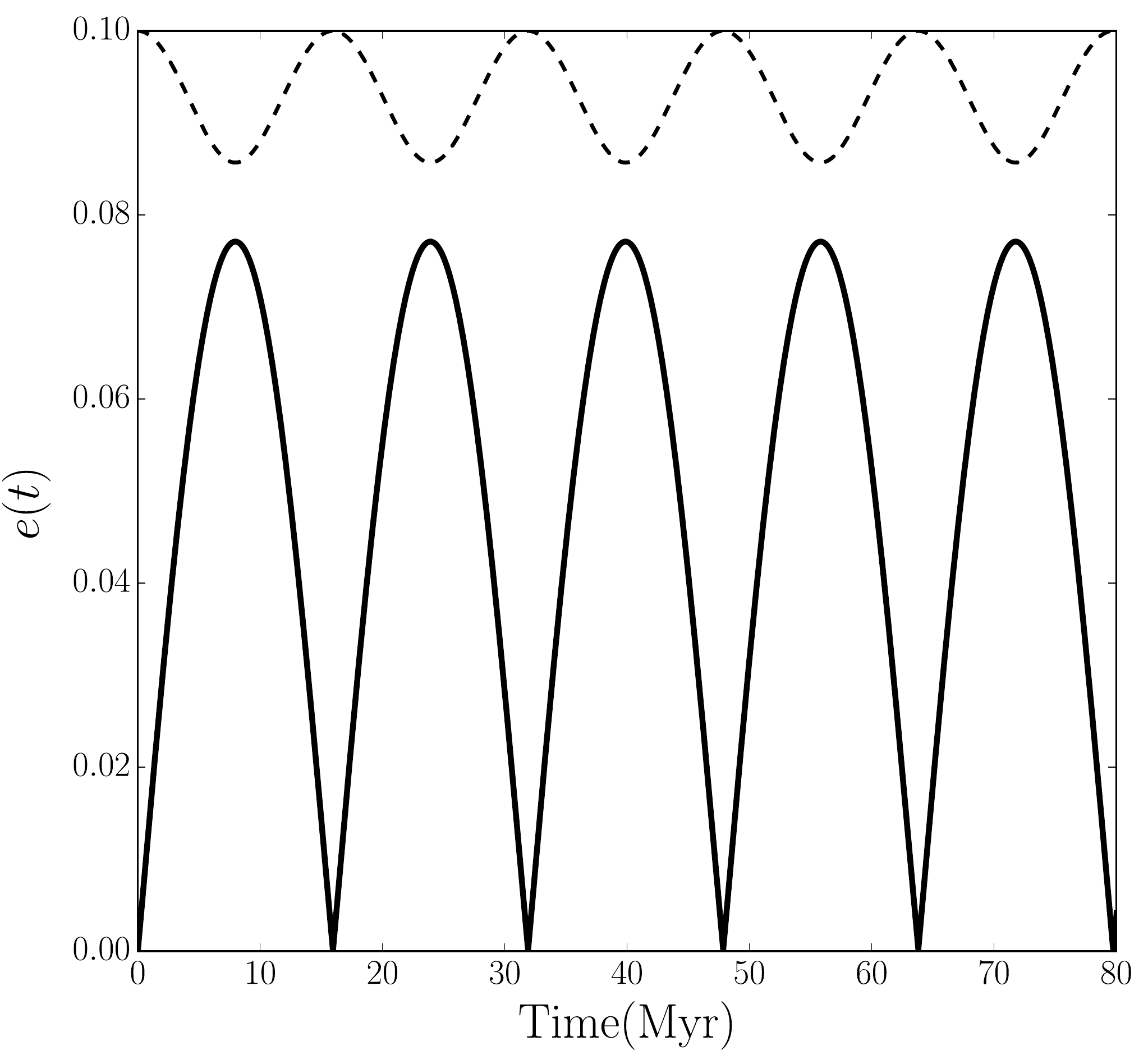}
	\includegraphics[width=0.42\linewidth]{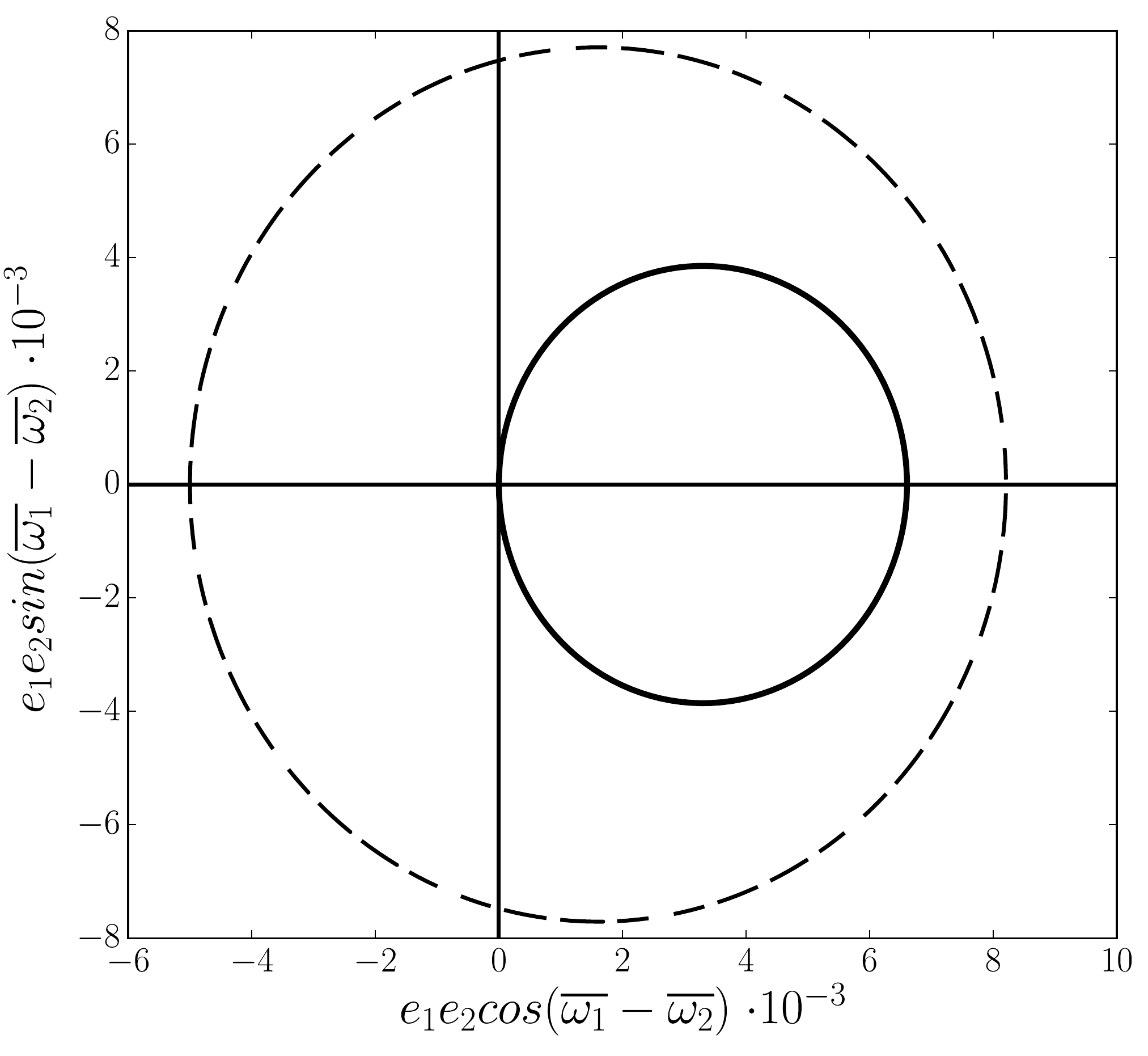}
	\caption{{\textit{(left)}} The evolution of eccentricity between an inner planet started on a circular orbit, $M_1$  (solid line) secularly interacting with a planet, $M_2$  (dotted line) of initial eccentricity $e_2(0)$ = 0.1. Each planet has 5M$_\oplus$ and placed at 1 and 5au respectively. The oscillation in the eccentricity of both planets, assuming that both eccentricities and inclinations are small, can be determined through a 2$\times$2 eigenfrequency problem via application of Laplace - Lagrange theory. The timescale of the precession in eccentricity is 2$\pi$/$g_1-g_2$, where $g_1,g_2$ are the two eigenfrequencies of the system in radians per year. {\textit{(right)}} Evolution of $z_1z_2^*$, where both complex eccentricities are given in eq \eqref{eq:secsol}, and $z_2^*$ is the complex conjugate of $z_2$. Such evolution for the example planets given in the left panel precesses around the solid circle. The dotted circle corresponds to when the planet on the initially circular orbit is initialised with an eccentricity of $e_1(0)$ = 0.05. The pericentres of the two planets are started 180$^\circ$ apart. Evolution is similar to the previous case, with the effect of the extra initial eccentricity component $e_1(0) \neq 0$ shifting the radius and position of the circle.}
	\label{fig:zevo}
\end{figure*}

\subsection{Application}
\label{subsec:sum}
For the purpose of this work we assume that eccentricities and inclinations are small, allowing for the direct application of Laplace - Lagrange theory to measure the eccentricity evolution of secularly interacting planets. We also assume that at high eccentricities this theory can still be applied as an approximation, which we quantify later in this paper through comparisons with N-body simulations. Finally, as eccentricities are assumed to be small we treat secular resonances as occurring at fixed locations and ignore any inclusion of secular chaotic interactions.   
It is possible to include MMRs in secular theory (\cite{1989A&A...221..348M}; \cite{1999MNRAS.303..806C}; \cite{2012ApJ...745..143A}), however to simplify the discussion we assume that planetary orbits are well spaced and away from MMRs. Thus we note that our results may not capture all of the relevant dynamics when considering planets with semi-major axis ratios defined by eq \eqref{eq:aMMR}. We can also assume that motion may be chaotic, and the system is unstable inside the region given in eq (\ref{eq:chaoswid}). 
If the separation of the pericentre and apocentre of neighbouring objects is within a distance of 5$R_H$, we also assume a scattering event can occur and the system is unstable. We use $R_H$ in the low eccentricity limit defined by eq \eqref{eq:mutualhill} for simplicity. We note that assumed scattering events are the more stringent constraint on how close planets can orbit when compared with the region of resonant overlap given in eq (\ref{eq:chaoswid}), as this is expected to only go out to $\sim$3 Hill radii (\cite{2000ApJ...534..428I}; \cite{2009Icar..199..197K}).

\section{Generalised 2-body Interactions}
\label{sec:2bod}

 Initially we consider the eccentricity evolution of two planets interacting via secular perturbations only, with the assumptions stated in \S\ref{subsec:sum}.
 These planets have masses and semi-major axes of $M_1$, $M_2$ and $a_1$, $a_2$ respectively, where $M_1$ is initialised on a circular orbit interior to $M_2$, which is on an eccentric orbit with an initial eccentricity $e_2(0)$. The respective initial longitudes of pericentre are $\varpi_1(0)$ and $\varpi_2(0)$.

 We propose various substitutions to the form of eq (\ref{eq:zevo}) to simplify the secular solution for the evolution of eccentricity. We give the ratio of the Laplace coefficients (eq (\ref{eq:laplace})) as the variable $f$ = $b^{(1)}_{3/2}(\alpha)${\LARGE{/}}$b^{(2)}_{3/2}(\alpha)$, where $\alpha = a_1/a_2$ and introduce the variable $L_i = M_ia_i^{1/2}$ for a planet of mass $M_i$ and semi-major axis $a_i$, which scales with orbital angular momentum for low eccentricity. The eccentricity solution for each planet ($z_1$, $z_2$) is then explicitly given by

\begin{equation}
\begin{aligned}
z_1(t) = \left(\frac{e_2(0)}{2y}\right)e^{i\varpi_2(0)}\left[e^{ig_2t} - e^{ig_1t}\right] ,  \\
z_2(t) = \left(\frac{e_2(0)}{2y}\right)e^{i\varpi_2(0)}\left[(y-x)e^{ig_1t} - (x+y)e^{ig_2t}\right],  
\label{eq:secsol}
\end{aligned}
\vspace{0.5cm}
\end{equation}
where $x = (f/2)(1 - L_1/L_2)$ and $y = \sqrt{x^2 + L_1/L_2}$. The associated eigenfrequencies are given by
\begin{equation}
\begin{aligned}
g_1 = -(A^*/2)\left[f(L_1 + L_2) + \sqrt{f^2(L_1 - L_2)^2 +4L_1L_2}\right] ,  \\
g_2 = -(A^*/2)\left[f(L_1 + L_2) - \sqrt{f^2(L_1 - L_2)^2 +4L_1L_2}\right],  
\label{eq:lamb}
\end{aligned}
\vspace{0.5cm}
\end{equation}
where $A^* = -\left(\pi a_1^{1/2} b^{(2)}_{3/2}(\alpha)\right)/\left(2\sqrt{M_c}a_2^{5/2}\right)$ in radians per year, for $a_1,a_2$ and $M_c$ in units of au and M$_\odot$ respectively. The left panel of Figure \ref{fig:zevo} shows the oscillation in eccentricity from eq \eqref{eq:secsol} for two example planets with $M_1$, $M_2$ of 5M$_\oplus$ each and $a_1$, $a_2$ of 1, 5au respectively with the planet on the initially eccentric orbit being initialised with $e_2(0)$ = 0.1. The timescale in years of a single period in this oscillation for both planets is given by $T = 2\pi/\left(g_1-g_2\right)$. 

Another way of visualising the evolution of $z_1(t)$ and $z_2(t)$ is to consider the evolution of $z_1z_2^{*}$ (where $z_2^*$ is the complex conjugate of $z_2$), which shows precession around a circle in an anti-clockwise direction (see solid circle in right panel of Figure \ref{fig:zevo}), highlighting the coupling of the secular solution that exists between the two planets.

The maximum in the eccentricity oscillation of the planet on the initially circular orbit, max[$e_1(t)$], as a result of the interaction with the planet on the initially eccentric orbit, as a function of its initial eccentricity $e_2(0)$, is given through the associated amplitude of $z_1(t)$ in eq \eqref{eq:secsol}
\begin{equation}
{\rm max}[e_{1}(t)/e_2(0)] = \left[\frac{L_1}{L_2} + \frac{1}{4}f^2\left(\frac{L_1}{L_2} - 1\right)^2\right]^{-1/2}.
\vspace*{5cm}
\label{eq:maxec}
\end{equation}

\begin{figure*}
	\centering
	\includegraphics[width=0.457\linewidth]{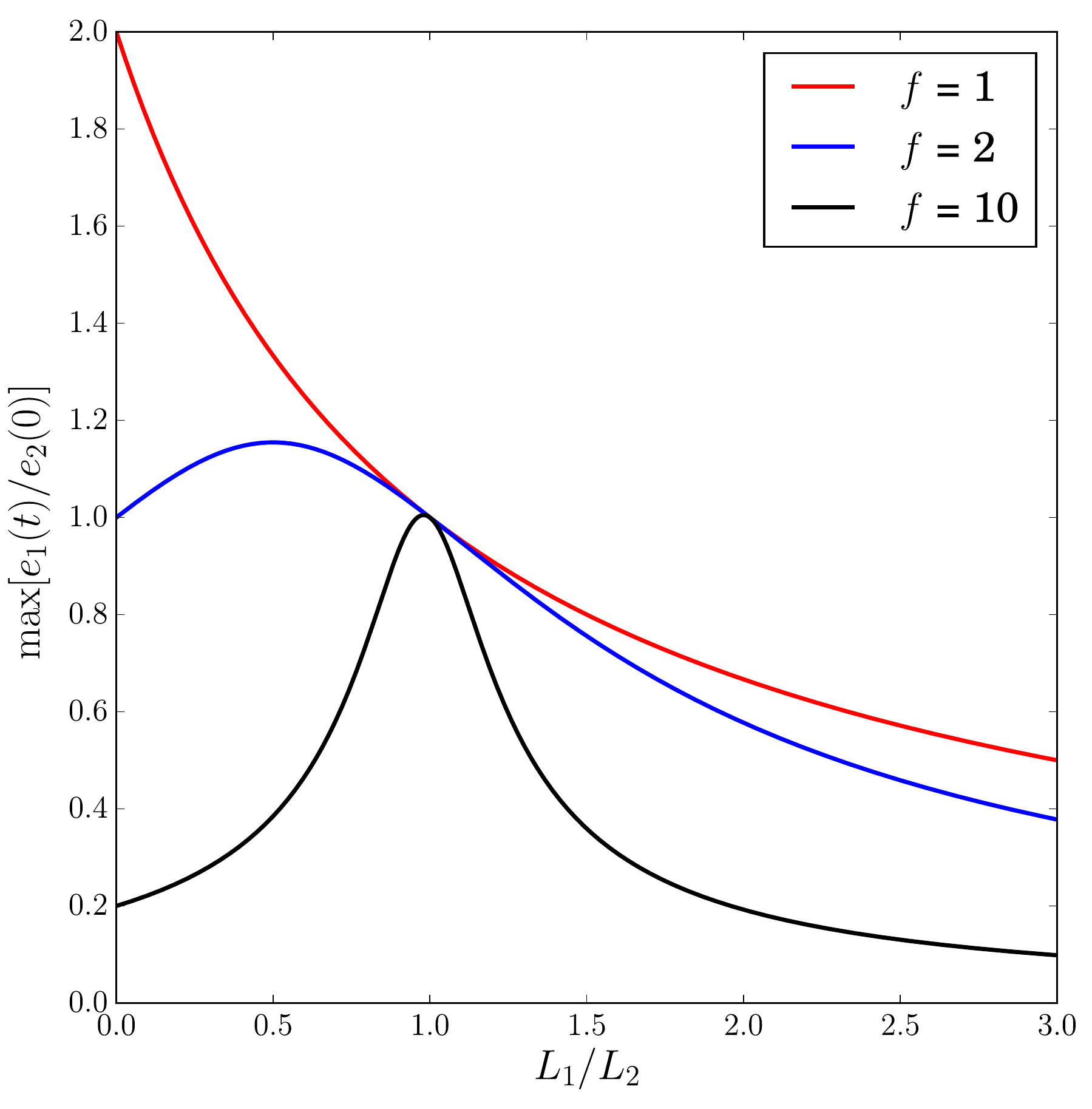}
	\includegraphics[width=0.51\linewidth]{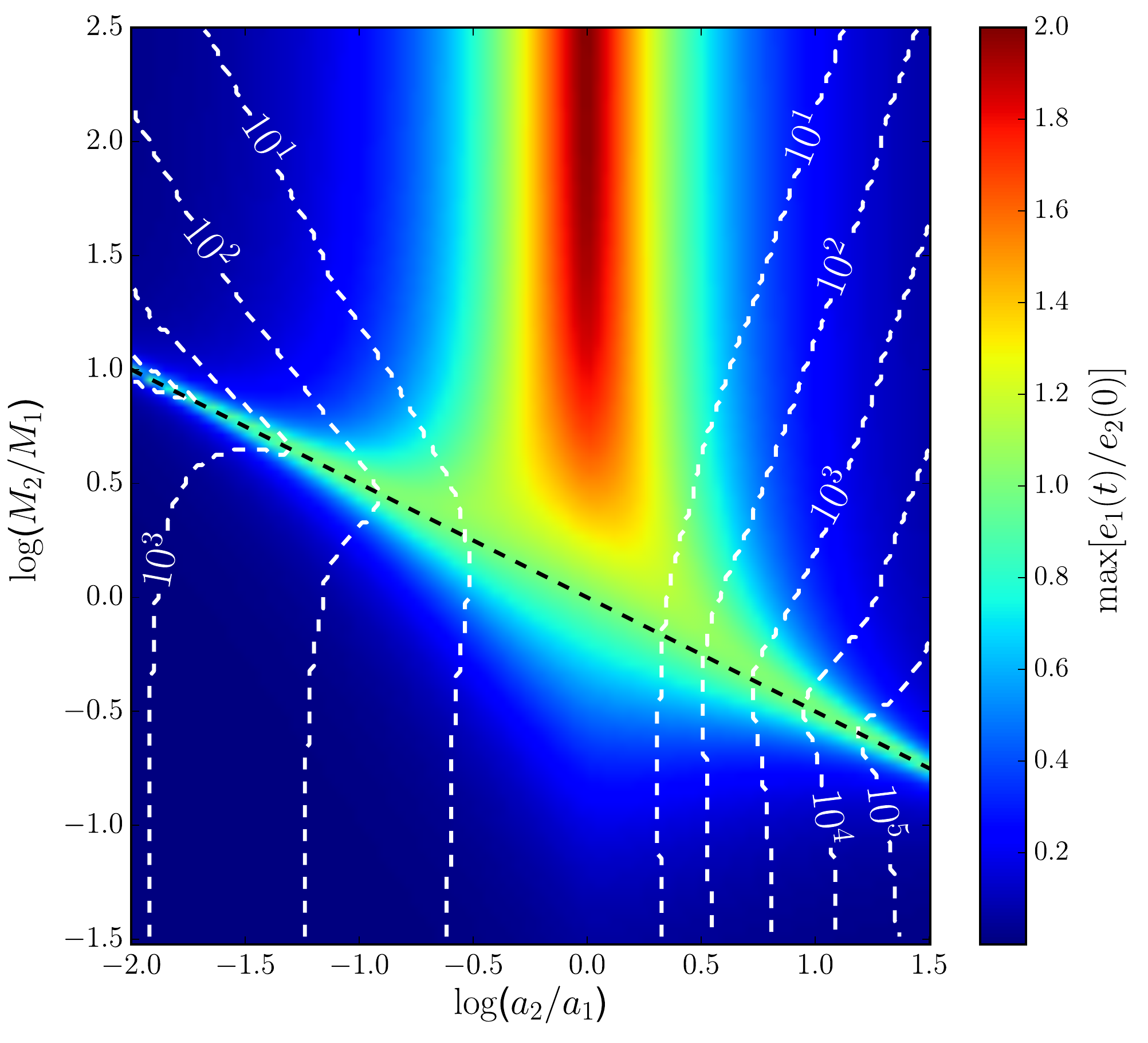}
	\caption{{\textit{(left)}} Maximum eccentricity of a planet on an initially circular orbit $M_1$ due to secular perturbations of a planet on an initially eccentric orbit $M_2$ as a function of $L_1/L_2$ (which is approximately the ratio of the angular momentum of the two bodies) for planets with different separations characterised by the parameter $f$. Regimes of equal orbital angular momentum between the two planets (in the low eccentricity limit) causes the maximum of the oscillation in eccentricity for the planet on the initially circular orbit to be equal to the initial eccentricity of the planet on the initially eccentric orbit. {\textit{(right)}} Colour scale gives the maximum eccentricity of a planet on an initially circular orbit due to secular perturbations of a planet on an initially eccentric orbit (as a function of its initial eccentricity $e_2(0)$) for a given mass and semi-major axis ratio between the two objects. The region of equal orbital angular momentum (in the low eccentricity limit) between the two planets is given by the black line, showing where the maximum eccentricity of the planet on the initially circular orbit is equal to $e_2(0)$. The white contours are values of $K$ in eq \eqref{eq:K}, which when multiplied by the orbital period of the planet on the initially circular orbit and its mass ratio with the central star ($M_c/M_1$), give the timescale of a single oscillation period in the eccentricity of both planets.} 
	\label{fig:maxe}  
\end{figure*}

Through inspection, it becomes evident that ${\rm max}[e_{1}(t)/e_2(0)]$ equals unity at points of $L_1$ = $L_2$, implying that the planet on the initially circular orbit will have a maximum eccentricity equal to the initial eccentricity of the planet on the initially eccentric orbit. Remembering our definition of $L_1,L_2$, it suggests that this will occur when the planets share an equal orbital angular momentum in the low eccentricity limit ($L_1 = L_2$). We show this in the left panel of Figure \ref{fig:maxe}, plotting eq \eqref{eq:maxec} as a function of $L_1/L_2$, for various values of $f$. 

The subtlety of varying $f$ becomes important when considering the value of $L_1/L_2$ that maximises eq \eqref{eq:maxec}. By taking associated derivatives, one finds this occurs at $L_1/L_2$ = $1-2/f^2$. For the $f \gg1$ regime, eq \eqref{eq:maxec} is indeed maximised when both bodies share equal orbital angular momentum (in the low eccentricity limit). When considering small $f$ values on the order of unity, eq \eqref{eq:maxec} has a maximised value where $e_1(t) > e_2(0)$ such that the maximum eccentricity of the planet on the initially circular orbit is greater than the initial eccentricity of the planet on the initially eccentric orbit. Realistically such $f$ values tend to limits where planets share small separations (1.0 > $\alpha \gtrsim 0.9$), which would most likely occur in regions where scattering or chaotic interactions can occur. Secular interactions of the most interest can therefore be assumed to be in the $f \gg 1$ regime.

When considering the limits of the left panel of Figure \ref{fig:maxe} they can be seen to agree with expectations, as for $L_1 \gg L_2$ (implying a case where $M_1 \gg M_2$), the planet on the initially circular orbit is massive enough to be unaffected by the secular perturbations of the planet on the initially eccentric orbit, hence it remains circular such that $e_1(t)$$\sim$0. The opposite case of $L_1 \ll L_2$ implies that the planet on the initially circular orbit instead tends to a limit where its behaviour is similar to that of a massless test particle ($M_1 \ll M_2$). One can evoke the well studied problem of a test particle in a secularly interacting system to explain this motion, which was touched upon in \S\ref{subsec:secular} and given in detail in \cite{1999ssd..book.....M}. In this case $z_1(t)$ precesses around a circle centred on the forced elements given by $z_2(0)/f$, resulting in an associated maximum eccentricity of ${\rm max} [e_1(t)] = 2e_2(0)/f$. 

We further demonstrate eq (\ref{eq:maxec}) by plotting it as a function of both the semi-major axis and mass ratio of both planets, $M_2/M_1$ and $a_2/a_1$ respectively, which we show in the right panel of Figure \ref{fig:maxe}. The maximum eccentricity induced in the planet on the initially circular orbit by the planet on the initially eccentric orbit, ${\rm max}[e_1(t)$/$e_2(0)]$, (eq \eqref{eq:maxec}, also shown by the vertical axis in the left panel of Figure \ref{fig:maxe}) is now represented by the colour scale. Our choice of axis allows for the planet on the initially eccentric orbit to be both interior and exterior to the planet on the initially circular orbit. 

It is clear that the maximum in the eccentricity oscillation of the planet on the initially circular orbit is equal to the initial eccentricity of the planet on the initially eccentric orbit (${\rm max}[e_1(t)$/$e_2(0)]$= 1 in eq \eqref{eq:maxec}) at regions of equal angular orbital momentum (in the low eccentricity limit) regardless of the $f$ value. This occurs along a straight line in the right panel of Figure \ref{fig:maxe}, which we denote in black. The low $f$ regime ($f$$\sim$1) can also be clearly identified in the right panel of Figure \ref{fig:maxe} at close separations between the planets ($\alpha$ $\sim$ 1). Here the maximum eccentricity of the planet on the initially circular orbit tends to the limit of ${\rm max} [e_1(t)] = 2e_2(0)/f$ from eq \eqref{eq:maxec}. As described before, some form of close encounter instability between the planets would most likely occur in this region. The high $f$ regime ($f \gg 1$) in the right panel of Figure \ref{fig:maxe}, where secular interactions are of most interest, refers to wider separations between the planets. The maximum eccentricity the planet on the initially circular orbit can have in this regime tends to the limit of $e_1(t) = e_2(0)$ from eq \eqref{eq:maxec}, which is traced by the black line (e.g. where the two planets have comparable orbital angular momentum).

As the right panel of Figure \ref{fig:maxe} is plotted in terms of ratios, the underlying shape of the maximum eccentricity of the planet on the initially circular orbit (e.g. the colour scale) will never change. This makes it completely generalised for direct application to similar two planet systems, observational or otherwise. 

We note that as the eccentricity of both planets oscillates on a timescale of $T = 2\pi/\left(g_1-g_2\right)$, the planet on the initially circular orbit will only be at a maximum in its eccentricity for a very small period of time. By substituting in eq \eqref{eq:lamb}, $T$ in years is explicitly given by 
\begin{equation}
\begin{aligned}
T = \frac{2}{\pi b^{(2)}_{3/2}(\alpha)}\left(\frac{a_2}{a_1}\right)^{5/2}\left[f^2\left(1 - \frac{L_2}{L_1}\right)^2 + 4\frac{L_2}{L_1}\right]^{-1/2}P_1\frac{M_c}{M_1},
\end{aligned}
\vspace*{5cm}
\label{eq:timescale}
\end{equation}
which can be simplified to say that 
\begin{equation}
T = KP_1\frac{M_c}{M_1}
\label{eq:K}
\end{equation}
where $P_1$ is the orbital period of the planet on the initially circular orbit. The value $K$ in eq \eqref{eq:K} is therefore a function of constants and the mass and semi-major axis ratios of the two planets given in eq \eqref{eq:timescale}. When the planet on the initially eccentric orbit is interior to the planet on the initially circular orbit ($a_2 < a_1$), a substitution can be made to $K$ in eq \eqref{eq:K} to keep eq \eqref{eq:timescale} as a function of $P_1M_c/M_1$. This is done by switching $a_1$ with $a_2$, $M_1$ with $M_2$ and vice versa. The $(a_2/a_1)^{5/2}$ factor is also replaced by ($a_1M_1/a_2M_2$). We show different values of $K$ through the dashed white lines in the right panel of Figure \ref{fig:maxe}.

While we have discussed the case where one of the planets started on an initially circular orbit, the evolution is similar if that planet were started on an eccentric orbit. For example the right panel of Figure \ref{fig:zevo} shows the evolution of $z_1z_2^*$ when the planet previously started on the circular orbit is initialised with an eccentricity $e_1(0)$=0.05 (dotted circle). Evolution is still around a circle but with a shifted centre and radius. The timescale of the interaction given in eq \eqref{eq:timescale} remains unchanged, with a dependence purely on mass and semi-major axis. In the $f\gg1$ regime, maximum transfer of eccentricity still occurs when $L_1\approx L_2$, though the exact maximum depends on the initial difference in the longitude of pericentres.

Whether the transfer of eccentricity predicted by eq \eqref{eq:maxec} occurs in systems with a higher number of planets is unclear as the secular solution becomes more complex (e.g. \cite{2004AJ....128..869Z}). We first apply this generalised 2-body secular theory to HD38858 by introducing a hypothetical planet into the system. We also now include treatment for how close planets can orbit before an instability may occur, outlined in \S\ref{subsec:sum}. We then see specifically if eq \eqref{eq:maxec} is still valid in describing a 3-body problem through application to 61Vir interacting with a hypothetical planet in \S\ref{sec:61Vir}.

\section{2 body application to HD38858}

\label{sec:HD38858}
 HD38858 is a relatively close, 0.9M$_\odot$ solar type G4V located in the constellation of Orion at a distance of 15.18 $\pm$ 0.09pc (\cite{2009ApJ...705...89L}; \cite{2012MNRAS.424.1206W}; \cite{2013ApJ...771...40B}). Various age estimates exist from 200Myr (\cite{2011A&A...530A.138C}), to  2.32 -- 8.08Gyr (\cite{2007ApJS..168..297T}) and 9.3Gyr (\cite{2010A&A...512L...5S}). Notably, HD38858 is part of a tentative correlation between low mass (sub-Saturn) planets and the presence of a debris disk (\cite{2012MNRAS.424.1206W}). As shown in Figure \ref{fig:HD3861Vir} the HARPS survey detects a single Super -Earth mass planet, HD38858b, with {\textit{Msini}} = 12.0 $\pm$ 2.0M$_\oplus$ and semi-major axis of 0.642 $\pm$ 0.002au (\cite{2015MNRAS.449.3121K}; Marmier et al. in prep). Derivation of further orbital parameters are expected to be presented in Marmier et al. (in prep), as such we simply consider a case where HD38858b has a circular orbit. It should be noted that a HD38858b was first detected with HARPS to have a minimum mass of $\sim$30M$_\oplus$ and semi-major axis of 1au (\cite{2011arXiv1109.2497M}), however this signal has since been identified as stellar activity (see \cite{2015MNRAS.449.3121K} for further discussion). Herschel DEBRIS imaging resolves disk structure at 30 -- 200au at an inclination of 44$^\circ$ $\pm$ 5$^\circ$ (\cite{2015MNRAS.449.3121K}), supported by inferred structure at 102au from $70$$\mu$m MIPS infrared excess (\cite{2012MNRAS.424.1206W}) and lower resolution Spitzer images (\cite{2012AJ....144...45K}). Hereafter we assume this to be the inclination of the planetary system and so consider the mass of HD38858b to be 17M$_\oplus$.
  
\begin{figure*}
  	\centering
  	\includegraphics[width=0.47\linewidth]{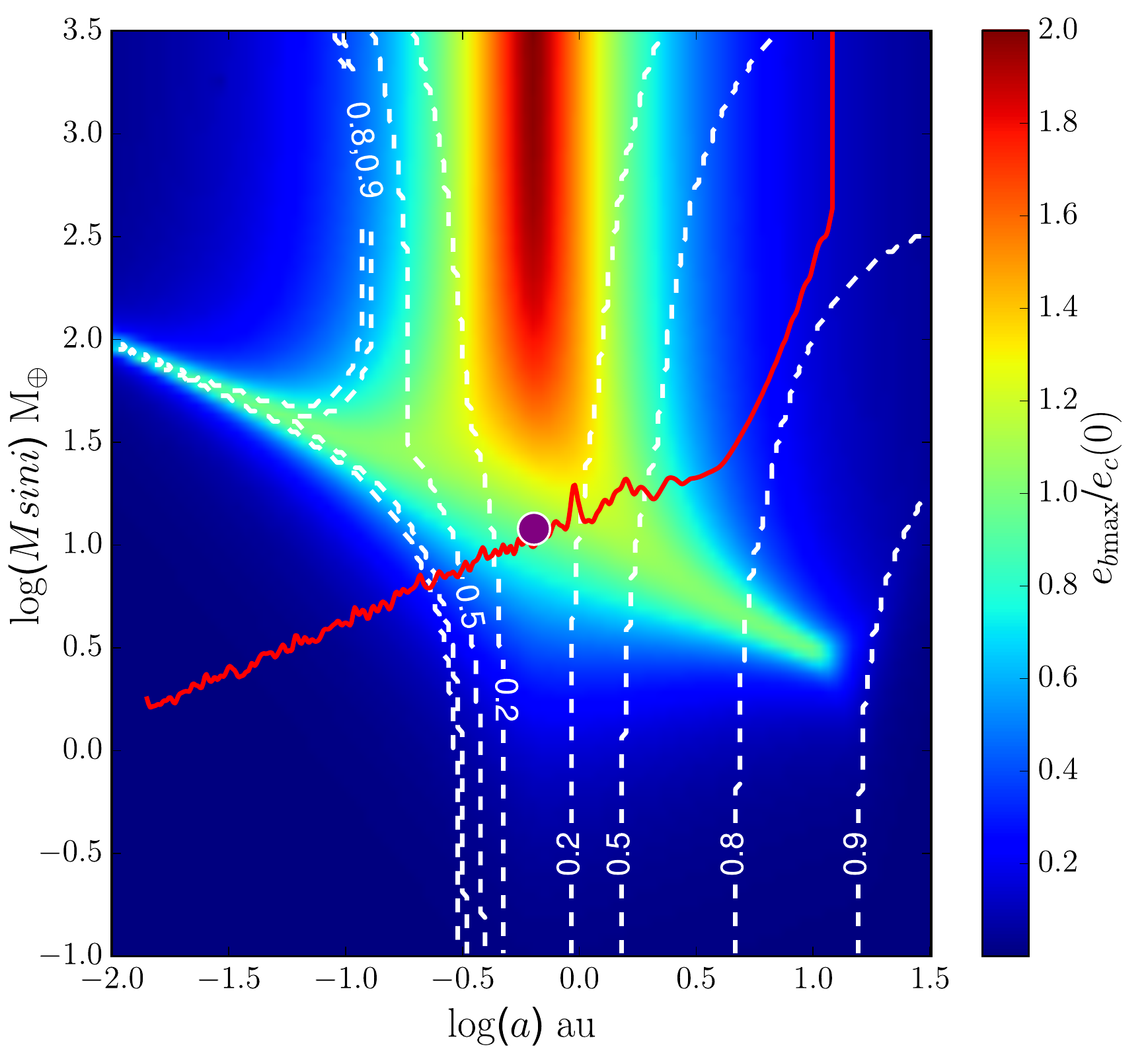}
  	\includegraphics[width=0.417\linewidth]{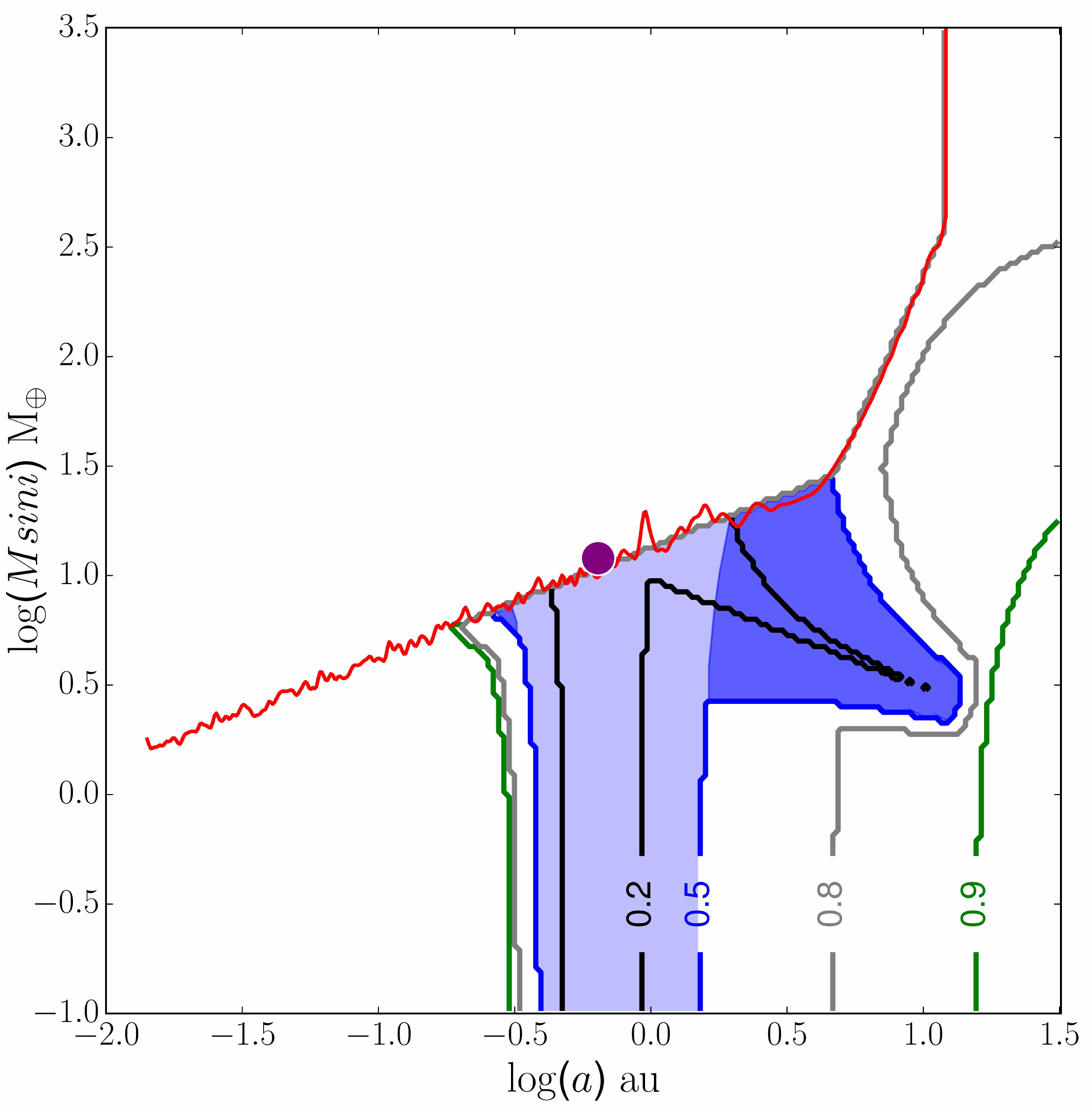}
  	\caption{{\textit{(left)}}: The maximum eccentricity of HD38858b ($e_{b\rm{max}}$) as a function of the initial eccentricity of a secularly perturbing hypothetical HD38858c ($e_c(0)$) of given {\textit{Msini}} and semi-major axis over 1Gyr. HD38858b is given by the purple circle with the HARPS RV sensitivity being given by the red line. White dashed contours enclose areas of assumed close encounters (<5$R_H$) for varying $e_c(0)$. {\textit{(right)}}: Contours represent constraints where a HD38858c with given initial eccentricity ($e_c(0)$) is unlikely to be present below the current HARPS RV sensitivity. This is due to either a close encounter with HD38858b being likely at some point in the secular evolution (light blue shaded region, explicitly for $e_c(0)$ = 0.5 for demonstration), or because HD38858b is secularly perturbed onto a significantly non-circular orbit ($e_{b\rm{max}}$>0.2, dark blue shaded region, explicitly for $e_c(0)$ = 0.5). Contour labels refer to different values of $e_c(0)$. For clarity, the light/dark blue shaded region is representative of where a HD38858c with an $e_c(0)$ = 0.5 would be unlikely to be present in HD38858 only, and is not fully representative of where a HD38858c would be unlikely to be present when $e_c(0)$ is changed.} 
  	\label{fig:HD38}
  \end{figure*}

 We introduce a hypothetical planet on an initially eccentric orbit to the system (HD38858c herein) and apply the secular interactions discussed in \S\ref{sec:2bod}. We largely ignore the presence of the disk and any additional interactions it would have with planets, using the inner edge as an outer constraint for the orbit of HD38858c only and assuming it to be coplanar with both planets for all time. HD38858c is therefore initialised with an {\textit{Msini}}, $M_c$, of 0.1M$_\oplus$ -- 10M$_{\rm{J}}$, semi-major axis, $a_c$, of 0.01 -- 30au and eccentricity, $e_c(0)$, of 0.1 - 0.9. 
 We refer to the maximum eccentricity of HD38858b due to its secular interaction with a given HD38858c as $e_{b\rm{max}}$. Due to the wide range of ages that exist for this system we simply calculate the secular interaction out to 1Gyr. We represent the ratio of $e_{b\rm{max}}/e_c(0)$ given by eq (\ref{eq:maxec}) as the colour scale in the left panel of Figure \ref{fig:HD38}. We plot masses in terms of $Msini$ to directly include the raw HARPS RV sensitivity of this system, which we give by the red line. The underlying shape of the colour scale is identical to that seen in the right panel of Figure \ref{fig:maxe} and therefore we make no further discussion of it here. However we note again that when HD38858b and HD38858c share equal orbital momentum (in the low eccentricity limit), the maximum in the eccentricity oscillation of HD38858b is equal to the initial eccentricity of HD38858c. For larger semi-major axes of HD38858c in Figure \ref{fig:HD38}, the timescale of the eccentricity oscillation increases according to eq \eqref{eq:timescale}. If this timescale is greater than 1Gyr, HD38858b does not have time to reach a maximum in the oscillation. This causes the difference between the colour scales of the right and left respective panels of Figure \ref{fig:maxe} and Figure \ref{fig:HD38} for a hypothetical HD38858c with $a_c$ >10au. 
 
 The maximum eccentricities shown in the left panel of Figure \ref{fig:HD38} are only valid in regions where the planets are always sufficiently separated for the effects of resonance overlap and close encounters between the planets to be negligible. The left panel of Figure \ref{fig:HD38} also shows the regions where such close encounters might be expected to occur,  that is, when at some point in the secular evolution the planets can possibly come within 5$R_H$ as discussed in \S\ref{subsec:sum}. The regions are outlined by contours representing the cases for different initial eccentricities of HD38858c. As noted in \S\ref{subsec:sum} this is always a more stringent constraint than the planets being close enough for resonance overlap. The size of this close encounter regime depends on the initial eccentricity of HD38858c and it can be assumed that inside this region the 2-planet system is unstable on a short timescale. 

  \begin{figure*}
  	\centering
  	\includegraphics[width=0.47\linewidth]{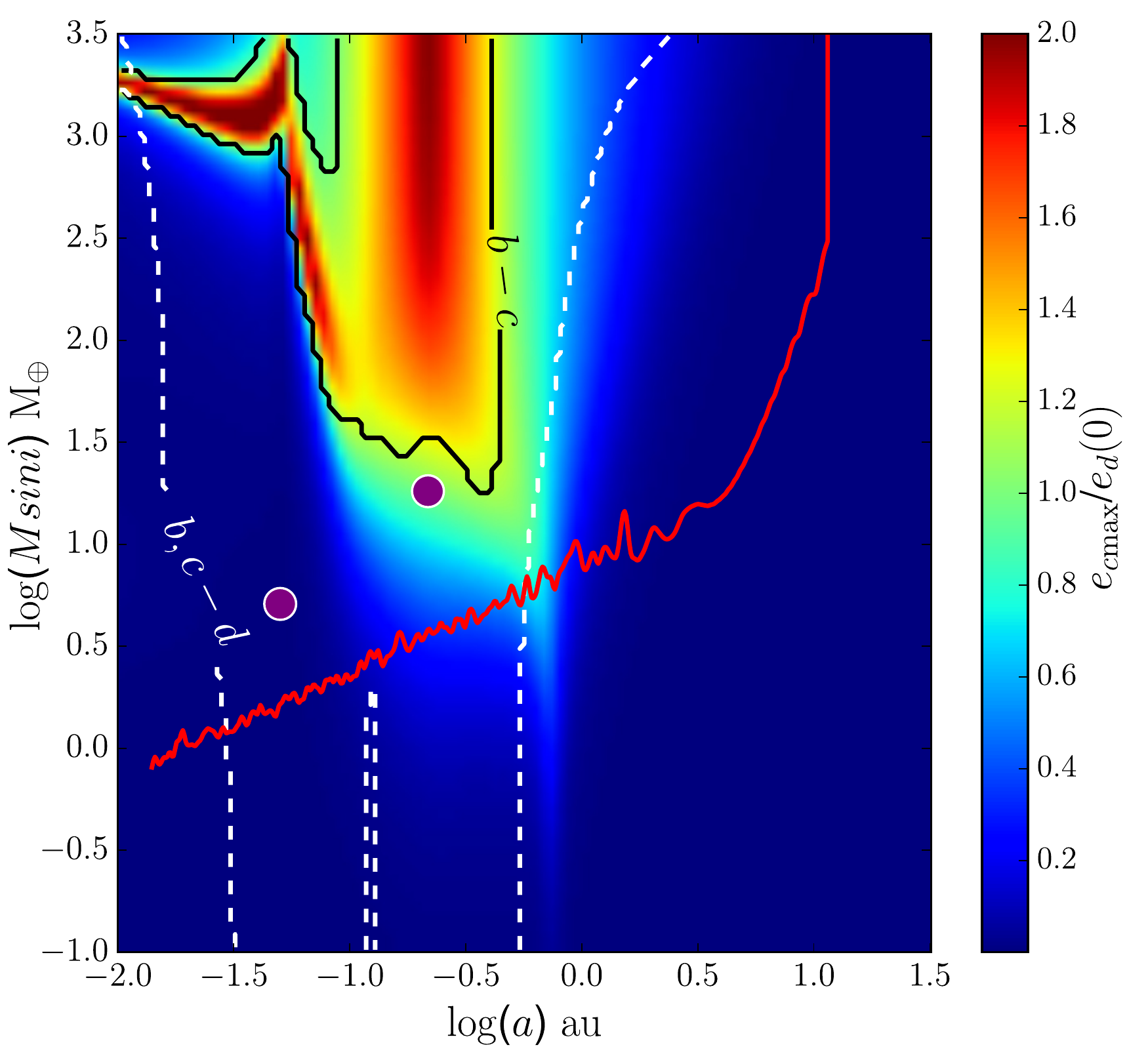}
  	\includegraphics[width=0.417\linewidth]{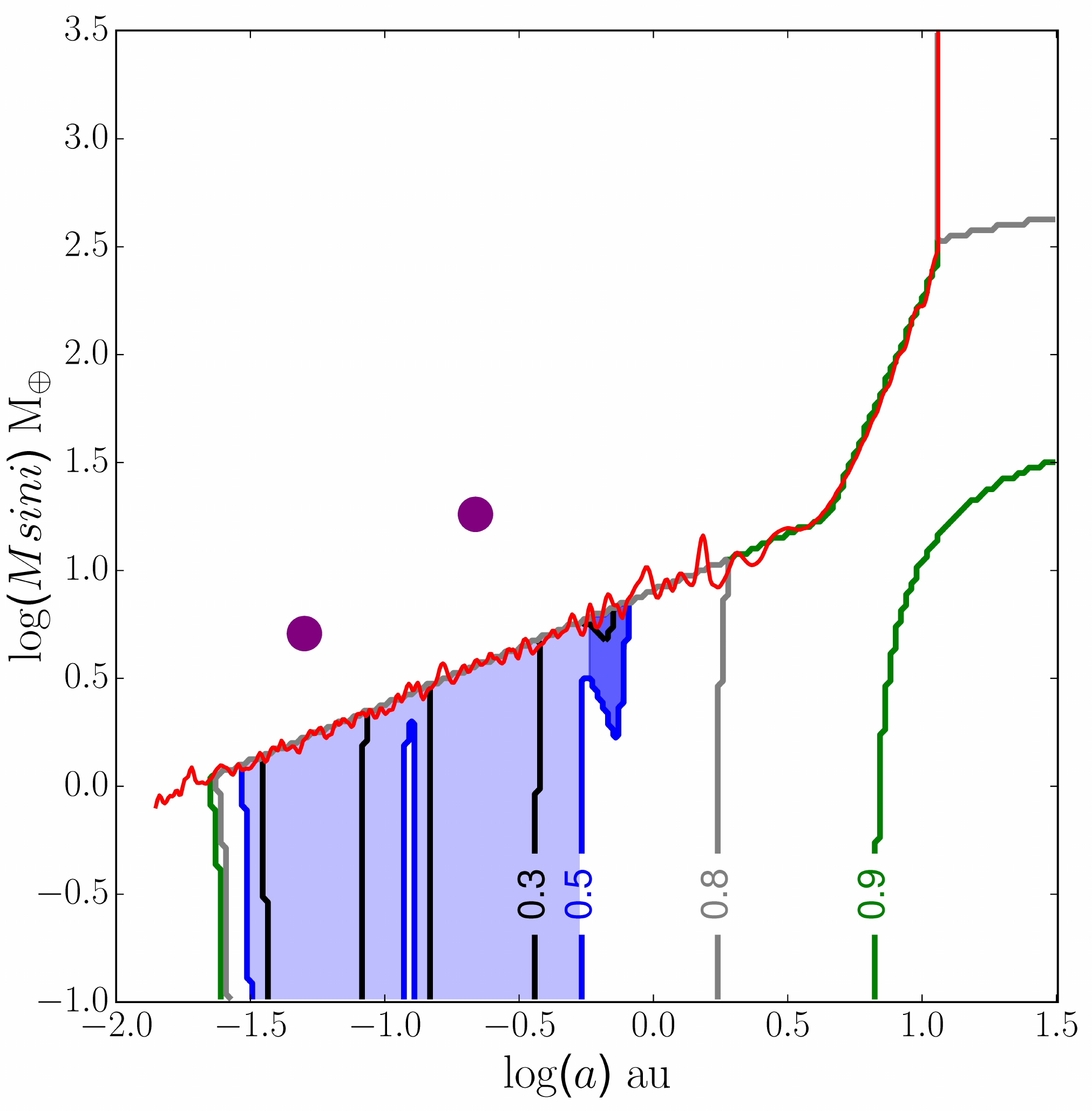}
  	\caption{{\textit{(left)}}: The maximum eccentricity of 61Vir-c ($e_{c\rm{max}}$) as a function of the initial eccentricity of a secularly perturbing hypothetical 61Vir-d ($e_d(0)$) of given mass and semi-major axis over 4.6Gyr. 61Vir-b/61Vir-c are given by the respective purple circles with the red line giving HARPS sensitivity. The white contour encloses a region where a close encounter would be likely between 61Vir-d with $e_d(0)$ = 0.5 and 61Vir-b or 61Vir-c. The black contour encloses a region inside which a 61Vir-d with $e_d(0)$ = 0.5 causes large enough perturbations in the eccentricity of 61Vir-b and 61Vir-c for a close encounter between the latter two bodies to be likely. {\textit{(right)}}: Constraints on where a 61Vir-d of a given $e_d(0)$, represented by each contour, is unlikely to be present in 61Vir below the HARPS RV sensitivity. This is either due to a close encounter between 61Vir-d and 61Vir-c or 61Vir-b being likely (light blue shaded region, explicitly for $e_d(0) = 0.5$ for demonstration, or because 61Vir-c has a maximum eccentricity above 0.2 as a result of the secular interaction with 61Vir-d (dark blue shaded region, explicitly for $e_d(0) = 0.5$). Contour labels refer to different values of $e_d(0)$. The light/dark blue shaded region is representative of where a 61Vir-d with an $e_d(0)$ = 0.5 would be unlikely to be present in 61Vir only, and is not fully representative of where a 61Vir-d would be unlikely to be present when $e_d(0)$ is changed.}
  	\label{fig:61Vir}
  \end{figure*}
 
  By noting the minimum threshold of HARPS sensitivity, a hypothetical HD38858c can exist in HD38858 without contradicting current observations whilst potentially inducing a large eccentricity in the observationally confirmed HD38858b. This becomes interesting when considering possible constraints on hypothetical planets in HD38858 which are not detectable via current HARPS sensitivity. If HD38858b was indeed measured by Marimer et al. to be on a roughly circular orbit, it would suggest that a hypothetical HD38858c may be unlikely to be present in regions where it induces a significant maximum eccentricity in HD38858b. However this does not mean that a HD38858c cannot be present in such regions, as potential eccentricity oscillations in HD38858b would be sinusoidal, thus its orbit may still appear roughly circular at a given moment in time. Similarly if the timescale of this oscillation is longer than the age of the system, HD38858bs orbit would potentially not have had sufficient time to have been perturbed onto a non-circular orbit. Thus, such an analysis only points to regions where a hypothetical HD38858c is unlikely to exist rather than being strictly ruled out.  
  
  In the right panel of Figure \ref{fig:HD38} we show the regions below the HARPS RV sensitivity where a hypothetical HD38858c is unlikely to exist, either because of a potential close encounter interaction with HD38858b at some point in the secular evolution, or because the secular interaction between the two planets causes HD38858b to have a maximum eccentricity, given by eq (\ref{eq:maxec}), that is significantly non-circular (judged to be when $e_{b\rm{max}}$ >0.2). Each contour corresponds to a different HD38858c initial eccentricity. For clarity, we explicitly shade in blue the region where a HD38858c with an $e_c(0)$ = 0.5 below the HARPS RV sensitivity is unlikely to exist for the sake of demonstration. The light blue part of this shaded region refers to where such a HD38858c is unlikely to exist due to a potential close encounter with HD38858b at some point in the secular evolution. The dark blue part of this shaded region shows the further constraint of where such a HD38858c is unlikely to exist solely because the maximum eccentricity it induces in HD38858b from the secular interaction is too large ($e_{b\rm{max}}$ >0.2), rather than because the two planets orbit close enough for a potential close encounter at some point in the secular evolution. This second constraint (e.g. the equivalent of the dark blue shaded region in the right panel of Figure \ref{fig:HD38} for a HD38858c with an $e_c(0)$ = 0.5) also exists for a HD38858c with an 0.2$\lesssim e_c(0) \lesssim$0.8. We note that the shaded light/dark blue region in the right panel of Figure \ref{fig:HD38} is representative of where a HD38858c that has an eccentricity $e_c(0)$ = 0.5 only is unlikely to be present in HD38858. This light/dark blue shaded region is therefore not fully representative of where a HD38858c with a different initial eccentricity ($e_c(0) \neq$ 0.5) would be unlikely to be present in HD38858.
  
  While the region where HD38858c is unlikely to exist below the HARPS RV sensitivity, solely due to its secular interactions causing HD38858b to have an $e_{b\rm{max}}$ >0.2 only, may be relatively small, it encapsulates an interesting area of the parameter space, namely a HD38858c with a mass between 3-10M$_\oplus$ and semi-major axis between 1-10au. We conclude therefore that it is useful to consider if significant eccentricities are induced between secularly interacting planets, not only in HD38858 but also in other single planetary systems. If this were the case it would add an extra constraint on the regions where a hypothetical planet is unlikely to be present, in addition to the perhaps more obvious interaction of where the planets may scatter/eject one another.

 \section{3-body application to 61Vir}
 \label{sec:61Vir}
 At 8.55 $\pm$ 0.02pc (\cite{2007A&A...474..653V}), 61Virginis (61Vir/HD115617) is the 8th closest G-type star (G5V) to the Sun with a mass of 0.88M$_\odot$ (\cite{2007MNRAS.374..664C}; \cite{2008A&A...487..373S}). Age estimates place it to be relatively old, in the region of 3 -- 11.5Gyr (\cite{2005ApJS..159..141V}; \cite{2007MNRAS.374..664C};  \cite{2007ApJS..168..297T}), with most recent estimates from gyrochronology placing it at 4.6 $\pm$ 0.9Gyr (\cite{2011ApJ...743...48W}; \cite{2012AJ....143..135V}). KECK/HIRES and HARPS RV measurements confirm two Super-Earth type planets: 61Vir-b and 61Vir-c, with an $Msini$ and semi-major axis of 5.1 $\pm$ 0.5M$_{\oplus}$, 18.2 $\pm$ 1.1M$_{\oplus}$ and 0.05au, 0.218au respectively, shown in Figure \ref{fig:HD3861Vir} along with the current HARPS sensitivity (\cite{2010ApJ...708.1366V}; \cite{2012MNRAS.424.1206W}). The full set of orbital elements for 61Vir-b and 61Vir-c derived by \cite{2010ApJ...708.1366V} are given in Table \ref{tab:61Vir}. \cite{2010ApJ...708.1366V} also detected a third planet with $Msini$ and semi-major axis of 22.9 $\pm$ 2.6M$_\oplus$ and 0.48au respectively. However much like the original detection of HD38858b, this was identified to be a stellar signal (\cite{2012MNRAS.424.1206W}). Herschel DEBRIS imaging has identified a debris disk at $\sim$30-100au with an inclination ($i$) of $\sim$77$^\circ$ (\cite{2012MNRAS.424.1206W}). We again assume this to be the inclination of the planetary system and so consider that 61Vir-b and 61Vir-c have a mass of 5.2M$_\oplus$ and 18.7M$_\oplus$ respectively.
 
 \indent We consider the 3-body secular problem of 61Vir-b and 61Vir-c interacting with a hypothetical planet on an eccentric orbit (called 61Vir-d herein). We calculate the maximum eccentricity induced in the outer planet, 61Vir-c, from 61Vir-d which we refer to as $e_{c\rm{max}}$ (with the effects on/from 61Vir-b still being included in our calculations). Any deviation from the 2-body interaction described by eq \eqref{eq:maxec} is therefore due to the presence of the third body (61Vir-b in this case). For simplicity we initialise 61Vir-b and 61Vir-c on circular orbits, since derived eccentricities are also consistent with $\sim$0 from Table \ref{tab:61Vir}, and assume all objects to be coplanar with the disk for all time. We also calculate the secular interaction out to 4.6Gyrs. We include 61Vir-d in the same way as HD38858c in \S\ref{sec:HD38858}, using the inner edge of the disk (30au) as an outer constraint only, initialising 61Vir-d with an $Msini$ (M$_d$) of 0.1M$_\oplus$ -- 10M$_{\rm{J}}$, semi-major axis ($a_d$) of 0.01 -- 30au and eccentricity ($e_d$) of 0.1 - 0.9. 
 
 The colour scale of the left panel of Figure \ref{fig:61Vir} shows the maximum eccentricity induced in 61Vir-c as a function of the initial eccentricity of 61Vir-d ($e_{c\rm{max}}/e_{d}(0)$) that has a given $Msini$ and semi-major axis. As the colour scale is a function of $e_{c\rm{max}}/e_{d}(0)$, it is independent of the initial eccentricity of 61Vir-d. Comparing the left panels of Figures \ref{fig:HD38} and \ref{fig:61Vir}, there is a clear difference between the 2 and 3-body secular interaction, as the maximum eccentricity of 61Vir-c is not equal to the initial eccentricity of 61Vir-d in regions where they share an equal orbital angular momentum (in the low eccentricity limit). Moreover outside of $\sim$1au in the left panel of Figure \ref{fig:61Vir}, $e_{c\rm{max}}$ becomes very small for all masses of 61Vir-d. As such, eq \eqref{eq:maxec} does not describe the maximum eccentricity 61Vir-c will have due to the secular perturbations of 61Vir-d. The presence of 61Vir-b can therefore reduce the effect of the secular perturbations of 61Vir-d on 61Vir-c, even for a 61Vir-d outside of $\sim$1au that is significantly eccentric and massive. We also note that the eccentricity of 61Vir-d is also unchanged from the interaction with 61Vir-b and 61Vir-c.   
 
 This 'stabilising' of 61Vir-c by 61Vir-b can also be inferred from where the parameters of 61Vir-d cause the eccentricities of 61Vir-b and 61Vir-c to become large enough, such that one of the planet pairs come within a distance that may lead to a potential close encounter at some point in the secular evolution (<5$R_H$, outlined in \S\ref{subsec:sum}). In the left panel of Figure \ref{fig:61Vir}, we show that 61Vir-d with an $e_d(0) = 0.5$, causes such a potential close encounter specifically between 61Vir-b and 61Vir-c (solid black contour), only when this 61Vir-d itself already has a large enough eccentricity to potentially experience a close encounter with either 61Vir-b or 61Vir-c at some point in the secular evolution (dotted white contour). This behaviour is true regardless of the initial eccentricity of 61Vir-d.

 Therefore we conclude that it is not possible for a highly eccentric, very massive 61Vir-d on a wide orbit below the HARPS RV sensitivity to induce a significant eccentricity in the inner planets through secular perturbations, causing them to potentially collide/eject each other at some point in the secular evolution. A caveat exists however that if a greater number of planets are included at much wider semi-major axes than we have considered here ($\sim$ 50-200au), then propagation of eccentricities through the planets may be expected, potentially causing large eccentricities in the inner planets (\cite{2004AJ....128..869Z}).

 The process that causes 61Vir-b to stabilise 61Vir-c against the secular perturbations of 61Vir-d can be understood by considering the timescales of the interactions between each of the planets. From eq \eqref{eq:timescale}, 61Vir-b and 61Vir-c secularly interacting in isolation produces an eccentricity oscillation in both planets that varies over $\sim$22kyrs. Adding a 61Vir-d at 10au with 10M$_\oplus$ imposes an additional eccentricity variation on the other planets that occurs on a timescale of $\sim$Gyrs. Any coherent increase in the eccentricity of 61Vir-c caused by the secular perturbations of 61Vir-d would therefore be affected by the shorter period interaction between 61Vir-b and 61Vir-c. We note that it is this effect that has recently been the focus of explaining the dearth of planets around short period ($P$ $\lesssim$ 7d) binaries (\cite{2015arXiv150602039H}; \cite{2015MNRAS.449.4221H}; \cite{2015PNAS..112.9264M}; \cite{2015MNRAS.453.3554M}). How low mass a '61Vir-b' equivalent object can be and still stabilise the system is considered in further application to HD38858 (\S\ref{sec:stable}).
  
  As was discussed in \S\ref{sec:HD38858} and shown explicitly for HD38858c in the right panel of Figure \ref{fig:HD38}, constraints can be placed on where 61Vir-d would be unlikely to be present below the HARPS RV sensitivity (below the red line in Figure \ref{fig:61Vir}). This occurs when a given 61Vir-d orbits within 5$R_H$ of 61Vir-b or 61Vir-c at some point in the secular evolution and is assumed to experience a potential close encounter discussed in \S\ref{subsec:sum}. This also applies where the secular perturbations of a given 61Vir-d causes 61Vir-c to have a maximum eccentricity significantly above the currently derived value such that $e_{c\rm{max}}>0.2$ (Table \ref{tab:61Vir} at limit of uncertainty).  We show regions where 61Vir-d is unlikely to be present in the right panel of Figure \ref{fig:61Vir}. Each contour corresponds to a different initial eccentricity of 61Vir-d. Also in a similar way to what was shown in the right panel of Figure \ref{fig:HD38}, we explicitly shade in blue where a 61Vir-d below the HARPS RV sensitivity with an eccentricity of $e_d(0)=0.5$ is unlikely to exist for the sake of demonstration. The light blue region is where such a 61Vir-d experiences a potential close encounter with 61Vir-b or 61Vir-c at some point in the secular evolution. The dark blue region is where such a 61Vir-d causes 61Vir-c to have a maximum eccentricity $e_{c\rm{max}}>0.2$ only, without the two planets orbiting close enough for a potential close encounter at some point in the secular evolution. As described for the right panel of Figure \ref{fig:HD38} in \S\ref{sec:HD38858} we note that the shaded light/dark blue region in the right panel of Figure \ref{fig:61Vir} is representative of where a 61Vir-d that has an eccentricity $e_d(0)$ = 0.5 only is unlikely to be present in 61Vir, and is included for demonstration. This light/dark blue shaded region is not fully representative of where a 61Vir-d with a different initial eccentricity ($e_d(0) \neq$ 0.5) would be unlikely to be present in 61Vir.

  Comparing the right panels of Figure \ref{fig:HD38} and \ref{fig:61Vir}, it is clear that the excluded parameter space for where a 61Vir-d is unlikely to exist, solely due to the secular interaction causing 61Vir-c to have a maximum eccentricity $e_{c\rm{max}}>0.2$, is much less than what was seen for the 2-body problem in \S\ref{sec:HD38858} for HD38858c (e.g. comparing the dark blue shaded regions in the right panels of Figure \ref{fig:HD38} and \ref{fig:61Vir}). This area of parameter space below the HARPS RV sensitivity only excludes a 61Vir-d with $0.3\lesssim e_d(0) \lesssim 0.6$, for a mass and semi-major axis of 2-7M$_\oplus$ at $\sim$0.5au. Why this region is so much smaller than that seen for HD38858c in the right panel of Figure \ref{fig:HD38} can be understood to be due to 61Vir-b reducing the influence of the secular perturbations of 61Vir-d on 61Vir-c. We conclude therefore that we can only place very limited constraints on where a 61Vir-d of given $Msini$, $a_d$ and $e_d(0)$ would be unlikely to be present in 61Vir, by calculating where it induces a maximum eccentricity in 61Vir-c above its currently derived value of $\sim$0.2, and is such that it does not orbit close enough to 61Vir-c for a potential close encounter at some point in the secular evolution. We also suggest that this is more generally applicable to other planetary systems with two confirmed planets in close proximity to each other.

  To summarize we conclude that a consideration of secular interactions is unlikely to place significant constraints on an eccentric wide orbit planet that is secularly interacting with a pair of inner planets on close circular orbits, assuming that the wide orbit planet is below HARPS RV sensitivity. That is in order to place constraints on massive planets on wide eccentric orbits in 61Vir and other 2-planet systems, one cannot use the orbits of the known planets. Other than directly detecting a wide orbit planet, we suggest that investigations into the structure of an outer debris disk would provide the best type of constraints on such wide orbit planets. Planets on eccentric orbits have been shown to cause unique ring structure (\cite{2015MNRAS.453.3329P}), sculpt the inner edge (\cite{2012MNRAS.419.3074M}) and define the radial density profile (\cite{2014MNRAS.443.2541P}). For example a 61Vir-d would only need to be at $\gtrsim$5au to secularly interact with material at the inner edge of the disk on a timescale less than the age of the system (4.6Gyr).

 \section{Stabilising in HD38858}
 \label{sec:stable}
 We discuss a possible application of using the stabilising interaction between planets outlined in \S\ref{sec:61Vir} to infer the presence of additional planets in a 2-body system. As such we return to the example of the known HD38858 planetary system secularly interacting with a hypothetical additional planet on an eccentric orbit (HD38858c) discussed in \S\ref{sec:HD38858}. Consider such a HD38858c in the left panel of Figure \ref{fig:HD38} that causes a significant maximum eccentricity in HD38858b ($e_{b\rm{max}}$>0.2) through secular perturbations described in \S\ref{sec:2bod}. Also assume that the two planets do not orbit within 5$R_H$ at any point in the secular evolution and as such, a close encounter event between them is unlikely as outlined in \S\ref{subsec:sum}. Finally consider that this HD38858c is on the threshold of the HARPS RV sensitivity in the right panel of Figure \ref{fig:HD38} (on the red line), and that it is detected after the addition of further/reanalysis of RV data. For example, if this HD38858c had an initial eccentricity of $e_c(0)$ = 0.5, then this can be visualised in the right panel of Figure \ref{fig:HD38} by considering a HD38858c in an area where the HARPS RV sensitivity (red line) and the dark blue shaded region meets. 
 
 As HD38858b is assumed to have an eccentricity of zero, the presence of this 'newly detected' HD38858c could infer either of the following: (1) HD38858b is not close to a maximum in eccentricity as a result of the secular interaction with HD38858c at the time of observations. (2) The 'newly detected' HD38858c does not actually have a large enough eccentricity to cause a significant eccentricity in HD38858b through secular perturbations ($e_{b\rm{max}}$ < 0.2, discussed in \S\ref{sec:HD38858} and shown in eq \eqref{eq:secsol}). (3) There is a third object in HD38858 which is currently undetected, but is stabilising HD38858b against the secular perturbations of HD38858c in a way discussed in \S\ref{sec:61Vir}. We discuss the case where the last point is true.
 
 We consider how massive and at what semi-major axis a third body (a hypothetical HD38858d) would have to be to reduce the effects of the secular perturbations of a HD38858c on HD38858b. For simplicity we assume that the hypothetical HD38858d is interior to HD38858b, on a circular orbit and is also on the limit of HARPS RV sensitivity. We also assume that a close encounter interaction between HD38858b and the internal hypothetical HD38858d (\S\ref{subsec:sum}) is unlikely at any point in the secular evolution. This represents an ideal scenario where this stabilising inner HD38858d could also be detected upon follow up studies of RV data. 
 
\begin{figure}
  	\includegraphics[width=0.95\linewidth]{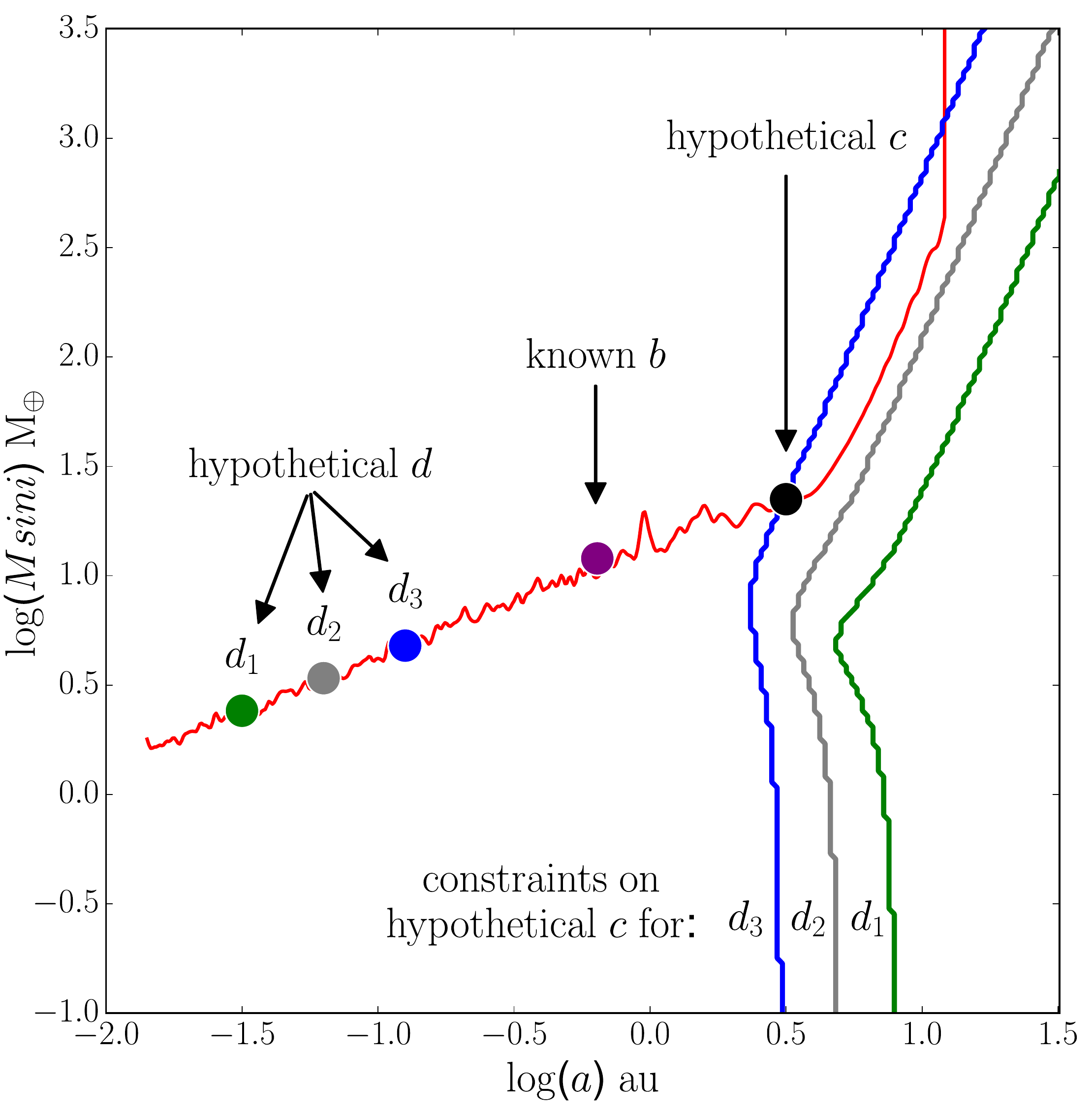}
  	\caption{HD38858b in the presence of an inner hypothetical planet $d$. The contours show where an inner $d$ reduces the secular perturbations of an outer hypothetical planet ($c$) on HD38858b by half. The black circle represents a specific outer hypothetical planet that should induce a large eccentricity in HD38858b from secular perturbations. If this planet were detected, it may infer the presence of planet interior to HD38858b located at the blue circle.}
 	\label{fig:HD38stab}
  \end{figure}

 Figure \ref{fig:HD38stab} represents HD38858b (Table \ref{tab:61Vir}) as the purple circle along with the HARPS RV sensitivity represented by the red line. We show three different example HD38858ds interior to HD38858b on circular orbits, which are on the limit of HARPS RV sensitivity with $Msini$ and semi-major axes of roughly 2.4, 3.4, 4.8M$_\oplus$ and 0.03, 0.06 and 0.13au, represented by the green, grey and blue circles respectively. We label each HD38858d as $d_1$, $d_2$ and $d_3$ respectively. The respectively coloured contours show where a HD38858c would need to exist to have the maximum eccentricity it induces in HD38858b, divided by its own eccentricity ($e_{b\rm{max}}/e_c(0)$), halved due to the presence of the inner HD38858d. At this point we assume that this HD38858d has stabilised HD38858b against the secular perturbations of a HD38858c. For example any HD38858c that lies along the green contour in Figure \ref{fig:HD38stab} will have the maximum eccentricity it induces in HD38858b divided by its own eccentricity, $e_{b\rm{max}}/e_c(0)$, reduced by half, due to the presence of the inner HD38858d located at the green circle ($d_1$). As the contours are ratios of $e_{b\rm{max}}/e_c(0)$ both before and after adding the internal HD38858d they are independent of the initial eccentricity of HD38858c.

 Consider for example a 'newly detected' HD38858c with an {\textit{Msini}} and semi-major axis of 22M$_\oplus$ and 3.2au respectively with an eccentricity of $e_c(0) =0.5$ (shown by the black circle in Figure \ref{fig:HD38stab}). Figure \ref{fig:HD38stab} shows that the maximum eccentricity it induces in HD38858b is reduced by half when a HD38858d internal to HD38858b on a circular orbit with 4.8M$_\oplus$ at 0.13au is present. Therefore if a HD38858c is detected at the black circle in Figure \ref{fig:HD38stab} with an eccentricity of $e_d(0) = 0.5$, then as this HD38858c should induce a significant eccentricity in HD38858b ($e_{b\rm{max}}$ >0.2, seen in the right panel of Figure \ref{fig:HD38}), this could also infer the presence of an additional planet, possibly located at the green circle on a circular orbit in Figure \ref{fig:HD38stab}.  
 
 This is obviously a very specific example dependant on many considerations. To our knowledge a two planet system where one planet is on a circular orbit and the other is on a significantly eccentric orbit has also yet to be discovered. However if future observations do detect such systems, we predict further planets might be inferred using the above method. We also note that it is possible that any stabilising effect could be due to multiple planets. This technique would still be useful in inferring that {\textit{something}} is causing a stabilising effect whether it is a single or many objects. 

 More generally we conclude that an inner planetary system made up of multiple planets is shielded from dynamical interactions from planets in the outer system. Moreover these inner planets do not need to be massive to have a protective effect on their neighbours. This reinforces the point that the full complement of planets in a system is needed to assess its dynamical state.

   \begin{figure*}
   	\centering
   	\includegraphics[width=0.48\linewidth]{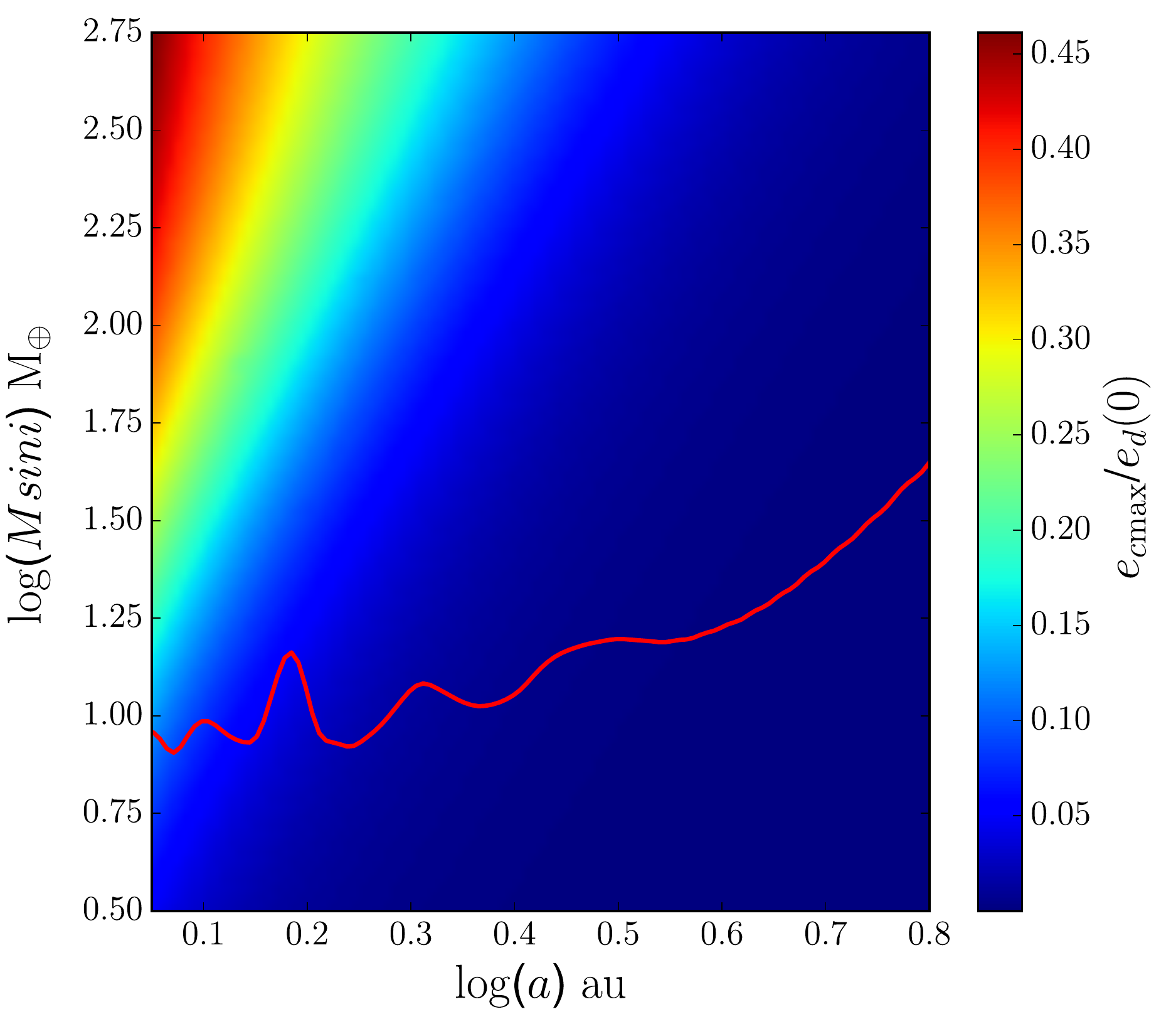}
   	\includegraphics[width=0.48\linewidth]{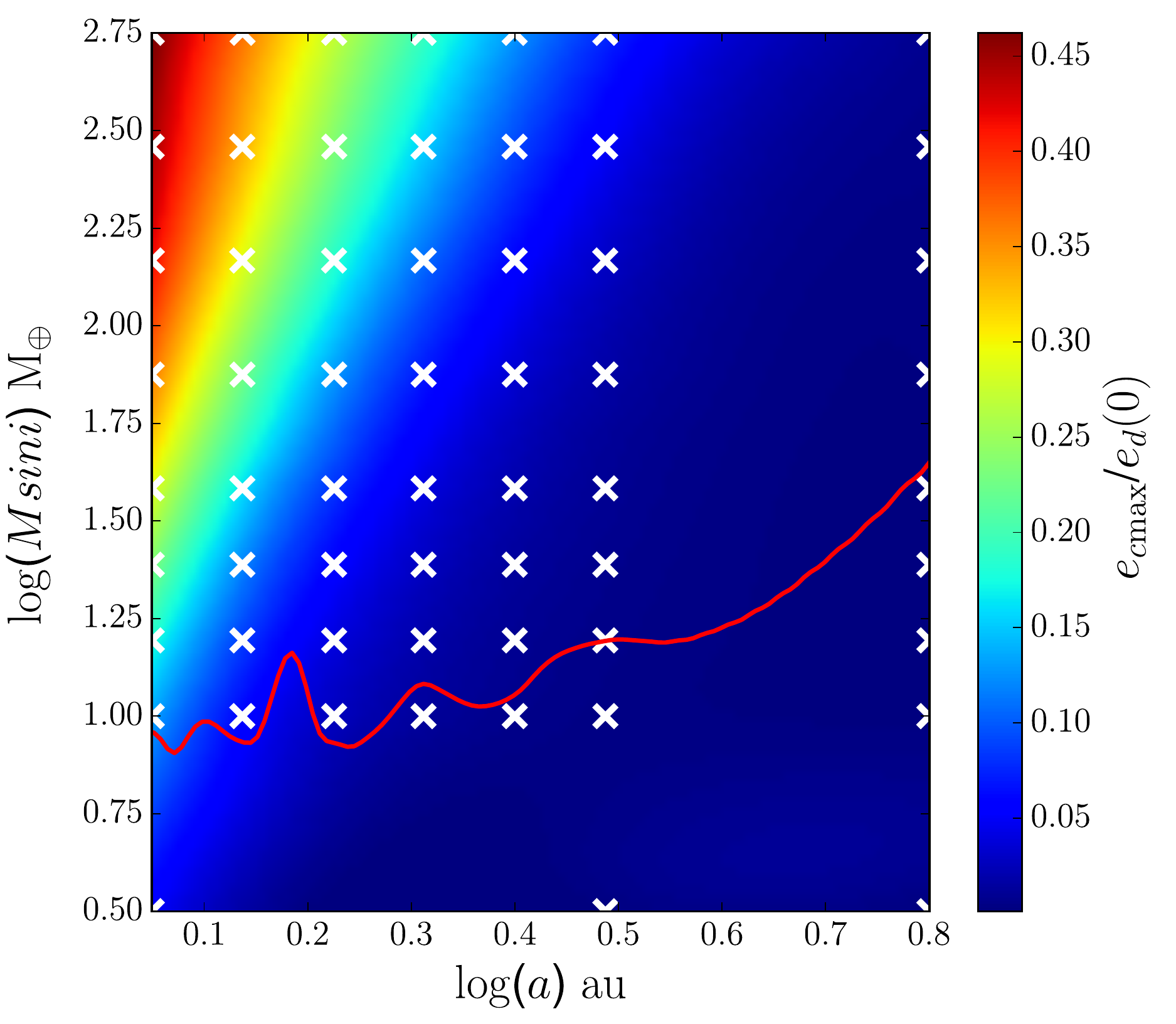}
   	\caption{{\textit{(left)}: The maximum eccentricity induced in 61Vir-c by 61Vir-d, divided by the initial eccentricity of 61Vir-d, ($e_{c\rm{max}}$/$e_d$) calculated by second order Laplace - Lagrange theory. This is simply a sub-region of Figure \ref{fig:61Vir}. The red line represents HARPS RV sensitivity. {\textit{(right)}}: $e_{c\rm{max}}$/$e_d$ calculated by N-body simulations. White crosses represent individual simulations for when 61Vir-d is initialised with an eccentricity of $e_d$ = 0.1. N-body simulations and second order Laplace - Lagrange theory predict the same $e_{c\rm{max}}$/$e_d$ in the low eccentricity limit as expected.}}
   	\label{fig:e0_1}
   \end{figure*}
   \begin{figure*}
   	\centering
   	\includegraphics[width=0.48\linewidth]{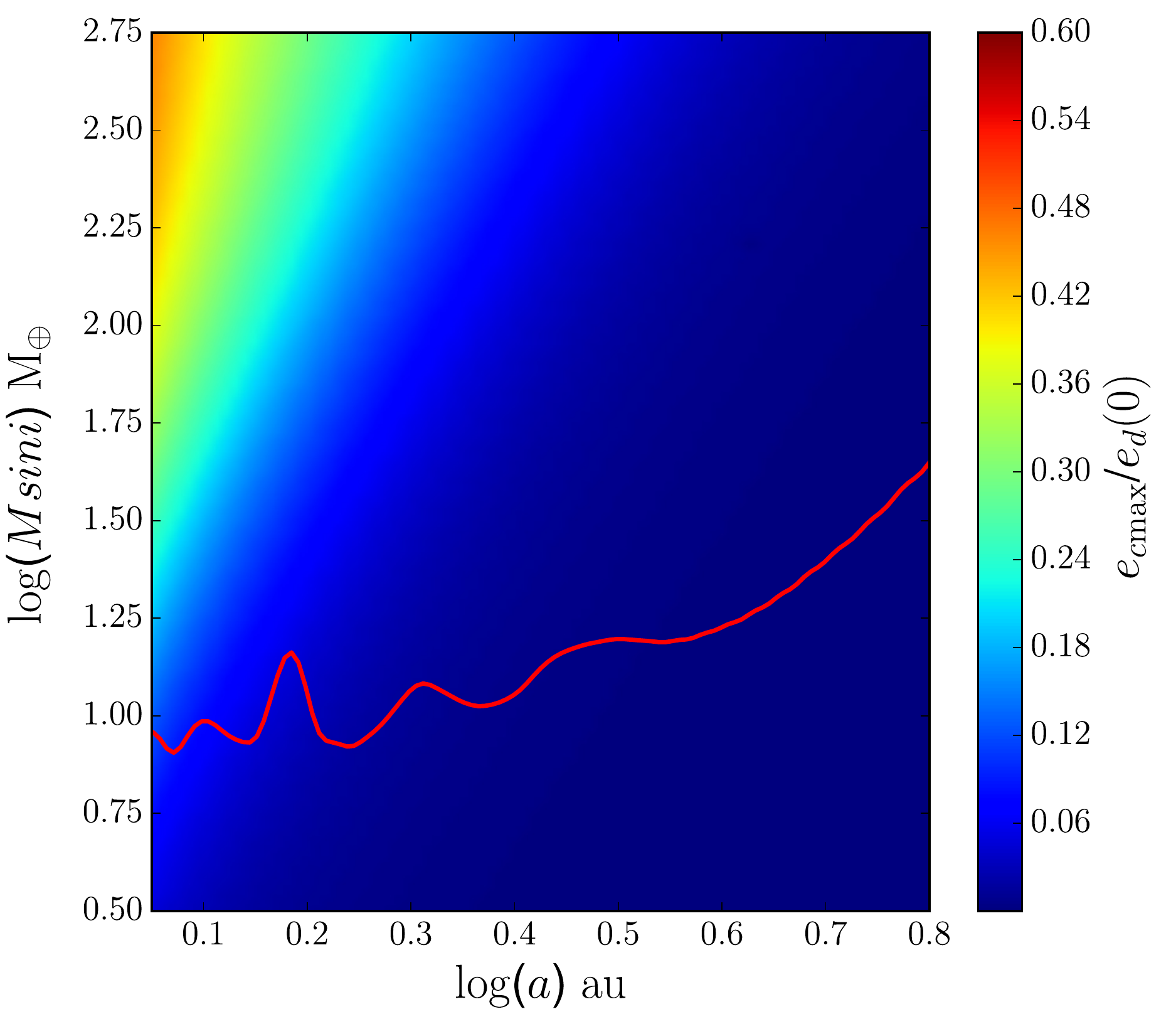}
   	\includegraphics[width=0.48\linewidth]{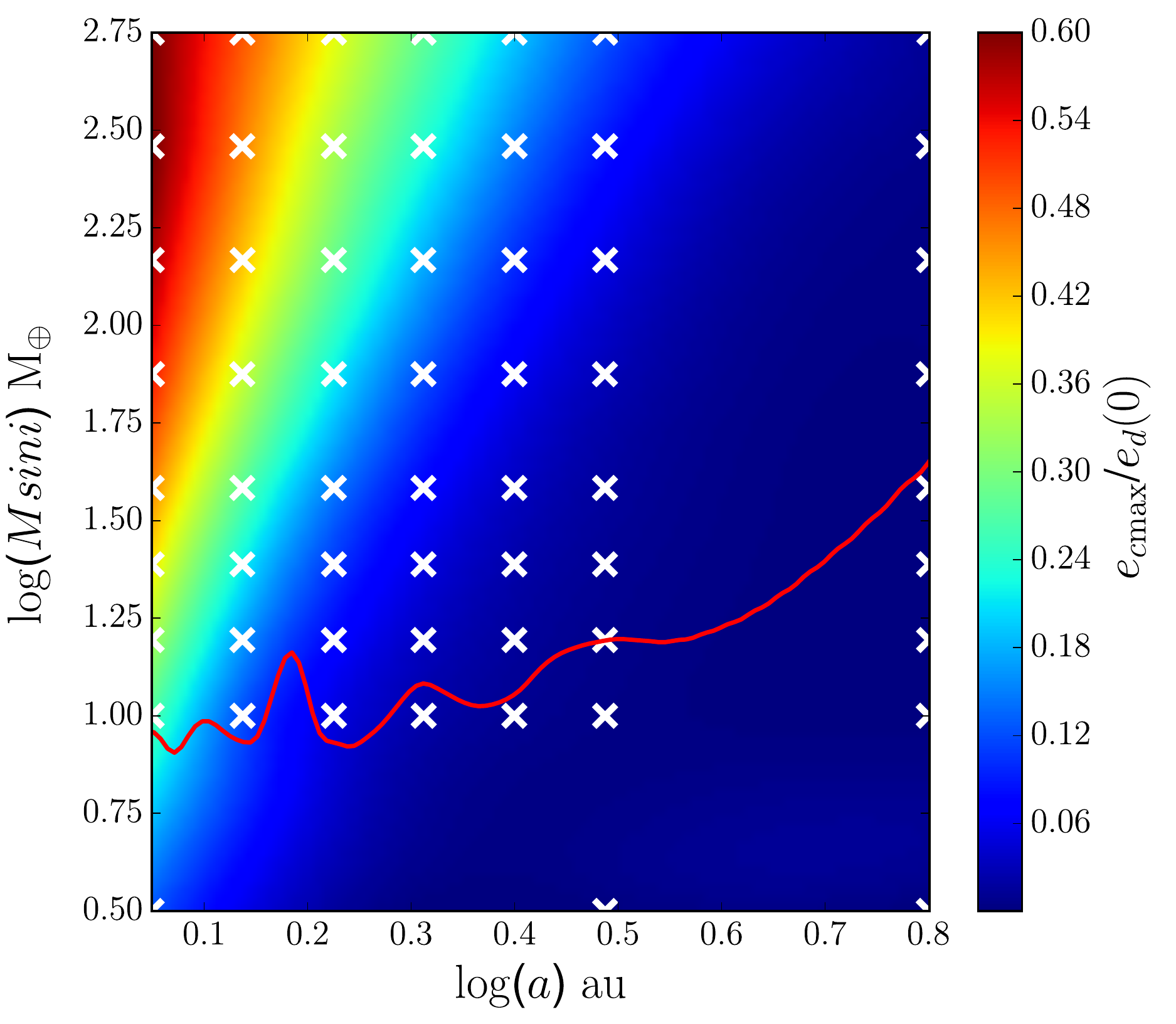}
   	\caption{{\textit{(left)}}: $e_{c\rm{max}}$/$e_d$ (see Figure \ref{fig:e0_1} caption for definition) calculated by second-order Laplace Lagrange theory (see Figure \ref{fig:61Vir}). The red line represents the HARPS RV sensitivity. {\textit{(right)}}: $e_{c\rm{max}}$/$e_d$ calculated by N-body simulations where the initial eccentricity of 61Vir-d is $e_d$ = 0.5. White crosses represent individual simulations. Comparing with the left panel, second order Laplace - Lagrange theory underestimates the maximum eccentricity that is induced in 61Vir-c by 61Vir-d from secular perturbations in the high eccentricity limit. However 61Vir-b can still be seen in simulations to reduce the effects of the secular perturbations of 61Vir-d on 61Vir-c.} 
   	\label{fig:e0_5}
   \end{figure*}
 
 \section{Limitations of Secular Theory}
 \label{sec:simulation}
 Thus far we have assumed eccentricities are small ($\lesssim0.2$) allowing for the application of second order Laplace - Lagrange theory (\S\ref{subsec:sum}) to calculate the evolution of eccentricity of secularly interacting planets. Once eccentricities go above $\sim$0.2 this theory begins to break down and higher order terms in the expansion of the disturbing function are required, as discussed in \S\ref{subsec:secular}. Assumptions for this work were also made in \S\ref{subsec:sum} that the effects of MMRs and long term secular chaotic interactions (e.g. \cite{2011ApJ...739...31L}) were negligible. We run N-body simulations to judge the validity of these assumptions, concentrating on application to 61Vir in \S\ref{sec:61Vir}, specifically to see if secular interactions are the dominant source of dynamical evolution in 61Vir and whether 61Vir-b actually damps the secular perturbations of a given 61Vir-d on 61Vir-c. 
 
 We use the N-body {\texttt{Mercury6-2}} code (\cite{1999MNRAS.304..793C}) and incorporate a hybrid integrator that switches from a more computationally efficient second order mixed symplectic algorithm, to a more detailed Bulirsch-Stoer integration when planets orbit close enough, assumed to be at <5 Hill radii for which the accuracy parameter is set to 10$^{-12}$. We set the timestep to 0.2 days, which is equal to $\sim$1/20 of the orbital period of 61Vir-b (Table \ref{tab:61Vir}). We neglect GR effects, however this would be expected to change the precession period of the eccentricity oscillation in eq \eqref{eq:zevo} rather than to have a significant effect on the maximum eccentricity (\cite{2007ApJ...661.1311V}; \cite{2013MNRAS.433.3190C}). 
 
 We run a grid of Mercury simulations for which 61Vir-d has an $Msini$ and semi-major axis of 10 -- 562M$_\oplus$ and 1.12 -- 6.31au respectively. This is assumed to be where the increase in the maximum eccentricity of 61Vir-c as a result of the secular interaction with 61Vir-d is of the most interest (see left panel of Figure \ref{fig:61Vir}). We run this grid twice, initialising 61Vir-d with an eccentricity of $e_d(0) = 0.1$ and $e_d(0) = 0.5$ to represent a low and high eccentricity regime respectively. We also assume that all planets are initially coplanar with the disk as described in \S\ref{sec:61Vir}. Each simulation is run to ten times the maximum secular period given by ${\rm{max}}(2\pi/\Delta g_i)$ where $\Delta g_i$ is the difference between a given pair of eigenfrequencies (implied by eq \eqref{eq:zevo}). Specifically this refers to run times of $\sim$10$^5$ - 10$^6$yrs. We also run a subset of simulations out to 3Gyr to look for the possibility of very long term chaotic secular effects (\cite{2011ApJ...739...31L}). For direct comparison with the left panel of Figure \ref{fig:61Vir}, the maximum eccentricity of 61Vir-c due to the secular interaction with 61Vir-d, divided by the initial eccentricity of 61Vir-d ($e_{c\rm{max}}/e_{d}(0)$) is calculated for each simulation.  
  
\begin{figure}
  	\includegraphics[width=1\linewidth]{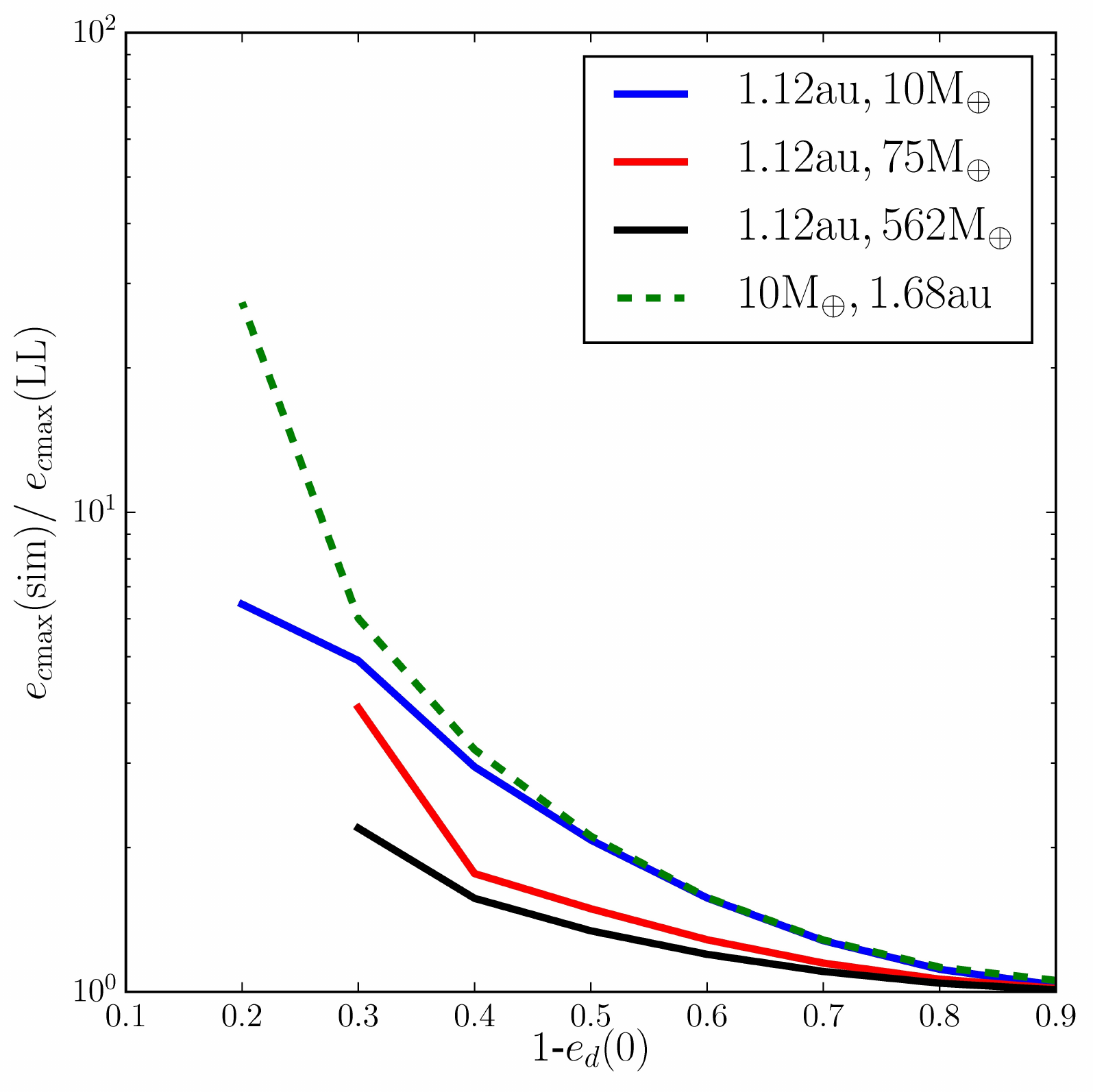}
  	\caption{The ratio of the maximum eccentricity induced in 61Vir-c by 61Vir-d due to secular interactions, divided by the initial eccentricity of 61Vir-d, $e_{c\rm{max}}$/$e_d$, calculated by N-body simulations to second order Laplace - Lagrange theory, as a function of the initial eccentricity of 61Vir-d. Eccentricity evolution predicted by second order Laplace - Lagrange theory shares a stronger agreement with simulations in the low eccentricity limit.}
 	\label{fig:sim_ana}
  \end{figure}

 Figure \ref{fig:e0_1} shows the comparison between $e_{c\rm{max}}/e_{d}(0)$ (given by the colour scale) for a 61Vir-d with a given $Msini$ and semi-major axis and initial eccentricity $e_d(0)$ = 0.1, calculated by second order Laplace - Lagrange theory (left panel of Figure \ref{fig:e0_1}, from \S\ref{subsec:secular}) and by Mercury simulations (right panel of Figure \ref{fig:e0_1}). Each individual simulation in the right panel of Figure \ref{fig:e0_1} is given by the white crosses, with the colour scale being a cubic interpolation over these points. Whereas, the interpolation was over 200$\times$200 grid in the left panel of Figure \ref{fig:sim_ana}, as were the previous Figures \ref{fig:maxe}, \ref{fig:HD38} and \ref{fig:61Vir}.  
 
 The left panel of Figure \ref{fig:e0_1} can be thought of a just a sub-region of the left panel of Figure \ref{fig:61Vir}. The values of $e_{c\rm{max}}/e_{d}(0)$ calculated by simulations and Laplace - Lagrange theory agree with each other to within a range of 6$\times$10$^{-3}$ (e.g. the maximum difference between values of the colour scale in the left and right panels of Figure \ref{fig:e0_1} is 6$\times$10$^{-3}$). This shows that second order Laplace - Lagrange theory is indeed a good predictor of secular evolution in the low eccentricity limit as expected. The right panel of Figure \ref{fig:e0_1} also shows that the reduced effect of the secular perturbations of 61Vir-d on 61Vir-c due to 61Vir-b described in \S\ref{sec:61Vir} is present in simulations, as a significant eccentricity is not induced in 61Vir-c when it has a similar orbital angular momentum to 61Vir-d. 

 Figure \ref{fig:e0_5} shows the same comparison between $e_{c\rm{max}}/e_{d}(0)$ calculated by second order Laplace - Lagrange theory (left panel of Figure \ref{fig:e0_5}) and simulations (right panel of Figure \ref{fig:e0_5}), however now setting the initial eccentricity of 61Vir-d to $e_d(0) = 0.5$. The values of $e_{c\rm{max}}/e_{d}(0)$ calculated by simulations and Laplace - Lagrange theory now only agree to within a range of 0.2 (e.g. the maximum difference between the colour scales of the left and right panels of Figure \ref{fig:e0_5} is 0.2). That is, simulations show that a significantly larger maximum eccentricity is induced in 61Vir-c by 61Vir-d, when compared with what is predicted by Laplace-Lagrange theory. This reflects the fact that higher order terms beyond those included in second order Laplace - Lagrange theory are required for large eccentricities as discussed in \S\ref{subsec:secular}. However as the same relative increase in the maximum eccentricity of 61Vir-c due to the secular interaction with 61Vir-d is seen in both the left and right panels of Figure \ref{fig:e0_5} (e.g. the gradient of the colour scales are similar), it suggests that 61Vir-b still reduces the effects of the secular perturbations of 61Vir-d on 61Vir-c in the high eccentricity limit. We conclude therefore that second order Laplace - Lagrange theory can be used as an approximation of the secular interaction between planets in 61Vir in the high eccentricity limit.  
 
 To highlight how much second order Laplace - Lagrange theory becomes an approximation of secular behaviour in the high eccentricity limit, we take a subset of our simulations and initialise 61Vir-d with an eccentricity $e_d(0) = 0.1 - 0.9$. This is done for a 61Vir-d with $Msini$ = 10, 75, 562M$_\oplus$ at a semi-major axis $a_d$ = 1.12au and for $a_d$ = 1.68au with $Msini$ = 10$_\oplus$. The ratio between the maximum eccentricity induced in 61Vir-c by 61Vir-d, divided by the initial eccentricity of 61Vir-d ($e_{c\rm{max}}/e_{d}(0)$) from simulations to second order Laplace - Lagrange theory is calculated for each of these $Msini$ and $a_d$ values of 61Vir-d. That is, a ratio of 1 means that simulations and second order Laplace - Lagrange theory predict an identical $e_{c\rm{max}}/e_{d}(0)$. Figure \ref{fig:sim_ana} plots this ratio against the initial eccentricity of 61Vir-d. The cut-off in some lines for high eccentricities occurs because a planet was ejected or scattered in our simulations. Figure \ref{fig:sim_ana} shows as the initial eccentricity of 61Vir-d is increased, $e_{c\rm{max}}/e_{d}(0)$ calculated by simulations becomes increasingly larger than what is predicted by second order Laplace - Lagrange theory, as expected from the discussion in \S\ref{subsec:secular}. 
 
 We finally note that for the area of parameter space of 61Vir-d in Figure \ref{fig:e0_1} and \ref{fig:e0_5}, MMRs seem to have a negligible effect, as simulations and second order Laplace - Lagrange theory show the same relative increase in the maximum eccentricity that is induced in 61Vir-c by 61Vir-d. This indicates that secular interactions between planets are the dominant source of dynamical evolution in 61Vir. However we note that a larger number of simulations would be required to confirm this, since our simulations only probed a limited number of semi-major axes and do not cover any significant MMRs. Our sub-set of simulations that were run out to long timescales ($\sim$3Gyr) also show no change in the maximum eccentricity of 61Vir-c as a result of the interaction with 61Vir-d, indicating that the effects of long term secular chaotic interactions are minimal.
 
\section{Summary and Conclusions}
\label{sec:conc}
We investigated how the dynamics of known exoplanet systems are affected by the presence of an extra hypothetical planet. We specifically focused on systems with 1 or 2 currently known planets, and considered the long term secular interaction with an additional hypothetical planet on an eccentric orbit. We subsequently investigated whether any constraints can be placed on such a hypothetical planet (in addition to those placed by HARPS RV measurements) by seeing if it were possible for this object to induce significant eccentricities in confirmed planets. We applied second order Laplace - Lagrange theory to calculate the eccentricity evolution of the planets due to their secular interaction, under the assumption that eccentricities remain small. 

We initially considered a generalised 2-planet system, where one planet was on an initially circular orbit and the other initially on an eccentric orbit. We showed that the maximum eccentricity of the planet on the initially circular orbit, due to the secular interaction, is equal to the initial eccentricity of the planet on the initially eccentric orbit, when both planets have a comparable orbital angular momentum.   

This generalised 2-planet secular interaction was then applied to the single planet system HD38858 in the presence of a hypothetical planet on an eccentric orbit. We showed that constraints can be placed on where this hypothetical planet would be unlikely to present below the HARPS RV sensitivity, solely due to the secular interaction causing an eccentricity in the known planet significantly above the low value observed. As potential eccentricity oscillations in the known planet would be sinusoidal however, its orbit may appear roughly circular at a given moment in time. Thus, this analysis places constraints on where a hypothetical planet is unlikely, rather than being strictly ruled out. These constraints apply to a hypothetical planet with an $Msini$ of 3-10M$_\oplus$, a semi-major axis between 1-10au, and an eccentricity in the range 0.2 - 0.8. We concluded therefore that secular interactions can be used to place significant constraints on hypothetical planets in single planetary systems, in addition to those placed by just considering where planets may scatter/eject on another. 

Whether the type of constraint from secular interactions can be applied to a hypothetical planet in a known 2-planet system was investigated through application to the 61Vir system in the presence of an additional hypothetical planet on an eccentric orbit. We showed that the maximum eccentricity of the outer known planet is no longer equal to the initial eccentricity of the hypothetical planet when the two planets have a comparable angular momentum. The inner known planet therefore reduces the effect of the secular perturbations of the hypothetical planet on the outer known planet. Constraints on where a hypothetical planet would be unlikely to be present in 61Vir below the HARPS RV sensitivity, from secular interactions alone causing an eccentricity in the outer known planet above the low value observed are limited. These constraints apply to a hypothetical planet with an $Msini$=2.7M$_\oplus$, a semi-major axis of $\sim$0.5au and an eccentricity in the range of 0.3 - 0.6. We concluded therefore that the orbits of known planets cannot be used to place significant constraints on massive planets on wide eccentric orbits in 61Vir and other 2-planet systems. We suggest that the structure of an outer debris disk may provide the best constraints on wide orbit hypothetical planets.

We investigated whether a hypothetical planet in the single planetary system HD38858, that is interior to the known planet can reduce the effects of the secular perturbations from an outer hypothetical planet. We showed that this can occur for the case where both the inner and the outer hypothetical planets are on the limit of HARPS RV sensitivity. This suggests that it may be possible to infer the presence of additional stabilising planets in systems with an eccentric outer planet and an inner planet on an otherwise suspiciously circular orbit. We concluded that inner planetary systems made up of multiple planets are shielded from dynamical perturbations from planets in the outer system. Moreover these inner planets do not need to be massive to have a protective effect on their neighbours. This reinforces the point that the full complement of planets in a system is needed to assess its dynamical state. 

Finally we show that for the case of the 61Vir system interacting with a hypothetical planet, N-body simulations also show that the inner known planet reduces the effects of the secular perturbations of a hypothetical planet on the outer known planet. We also show by comparisons with N-body simulations that second order Laplace - Lagrange theory can be used as an approximation of secular interactions in the high eccentricity limit. However in this limit an under prediction of the maximum eccentricities that are induced between secularly interacting planets is expected when calculated by second order Laplace - Lagrange theory.

\section*{Acknowledgements}
We thank Grant Kennedy for useful conversations regarding this work. We also thank the reviewer for helpful comments and suggestions. MJR acknowledges support of an STFC studentship and MCW acknowledges the support from the European Union through grant number 279973.




\bibliographystyle{mn2e}
\bibliography{example} 








\bsp	
\label{lastpage}
\end{document}